\documentclass[twocolumn,tighten]{aastex62}
\pdfoutput=1 
\usepackage{amsmath,amstext}
\usepackage[T1]{fontenc}
\usepackage{apjfonts} 
\usepackage[figure,figure*]{hypcap}
\usepackage{threeparttable}
\usepackage[htt]{hyphenat}
\usepackage{comment}

\begin{document}
\shorttitle{Small-scale clustering and lensing of BOSS}

\title{Joint analysis of small-scale galaxy clustering and galaxy--galaxy lensing from BOSS galaxies}

\author{Wenhao Gao}
\affiliation{Department of Astronomy, School of Physics and Astronomy, Shanghai Jiao Tong University, Shanghai 200240, China}

\author{Zhenjie Liu}
\affiliation{Department of Astronomy, School of Physics and Astronomy, Shanghai Jiao Tong University, Shanghai 200240, China}

\author{Zhongxu Zhai}
\affiliation{Department of Astronomy, School of Physics and Astronomy, Shanghai Jiao Tong University, Shanghai 200240, China}
\affiliation{Shanghai Key Laboratory for Particle Physics and Cosmology, Shanghai 200240, China}
\affiliation{Waterloo Center for Astrophysics, University of Waterloo, Waterloo, ON N2L 3G1, Canada}
\affiliation{Department of Physics and Astronomy, University of Waterloo, Waterloo, ON N2L 3G1, Canada}

\author{Jeremy L. Tinker}
\affiliation{Center for Cosmology and Particle Physics, Department of Physics, New York University, 726 Broadway, New York, NY 10003, USA}

\author{Jun Zhang}
\affiliation{Department of Astronomy, School of Physics and Astronomy, Shanghai Jiao Tong University, Shanghai 200240, China}
\affiliation{Shanghai Key Laboratory for Particle Physics and Cosmology, Shanghai 200240, China}

\author[0000-0002-5209-1173]{Arka Banerjee}
\affiliation{Department of Physics, Indian Institute of Science Education and Research,
Homi Bhabha Road, Pashan, Pune 411008, India}
\affiliation{Fermi National Accelerator Laboratory, Cosmic Physics Center, Batavia, IL 60510, USA}

\author{Joseph DeRose}
\affiliation{Lawrence Berkeley National Laboratory, 1 Cyclotron Road, Berkeley, CA 93720, USA}
\affiliation{Berkeley Center for Cosmological Physics, Department of Physics, UC Berkeley, CA 94720,
USA}

\author{Hong Guo}
\affiliation{Key Laboratory for Research in Galaxies and Cosmology, Shanghai Astronomical Observatory, Shanghai 200030, China}

\author[0000-0002-1200-0820]{Yao-Yuan~Mao}
\affiliation{Department of Physics and Astronomy, University of Utah, Salt Lake City, UT 84112, USA}


\author[0000-0001-8764-7103]{Kate Storey-Fisher}
\affiliation{Kavli Institute for Particle Astrophysics and Cosmology and Department of Physics, Stanford University, Stanford, CA 94305, USA}

\author[0000-0003-2229-011X]{Risa H. Wechsler}
\affiliation{Kavli Institute for Particle Astrophysics and Cosmology and Department of Physics, Stanford University, Stanford, CA 94305, USA}
\affiliation{Department of Particle Physics and Astrophysics, SLAC National Accelerator Laboratory, Stanford, CA 94305, USA}

\correspondingauthor{Zhongxu~Zhai}
\email{zhongxuzhai@sjtu.edu.cn}

\begin{abstract}
We present a joint analysis of galaxy clustering and galaxy--galaxy lensing measurements from BOSS galaxies using a simulation-based emulation method combined with a halo occupation distribution model. Our emulators are constructed with the Aemulus $\nu$ simulations, a suite of $w\nu$CDM $N$-body simulations with massive neutrinos as independent particle species. We combine small-scale analysis of clustering from $0.1h^{-1}$Mpc to $60.2~h^{-1}$Mpc and lensing from $1.7h^{-1}$Mpc to $60.2~h^{-1}$Mpc to perform cosmological constraints. We split the BOSS galaxies into three redshift bins to measure their clustering and employ galaxies from Dark Energy Camera Legacy Survey and Hyper Suprime-Cam as source galaxies to measure lensing separately. We find that the addition of lensing significantly improves the constraining power on $S_{8}=\sigma_8(\Omega_m/0.3)^{0.5}$, with a weak improvement for $f\sigma_{8}$. Our results of $f\sigma_{8}$ indicate tensions of around $1\sim4\sigma$ below the results of CMB observations of Planck. For $S_{8}$, our results are also lower than Planck, and the tension can be mitigated 
when considering possible systematics in lensing measurement. 
As a byproduct, our analysis prefers a non-zero neutrino mass but without strong significance, with the constraining power dominated by the clustering. Given the accuracy and precision of our model and the observational data, it is anticipated that larger and higher-quality spectroscopic datasets will improve the constraints on this fundamental property in the near future. 
\end{abstract}


\section{Introduction}

The large-scale structure of the universe contains a wealth of cosmological information and serves as one of the primary tools for probing cosmic evolution and constraining cosmological parameters. The spatial distribution of galaxies, as revealed by large-scale galaxy surveys, acts as a direct tracer of the underlying matter. Over the past few decades, a series of galaxy redshift surveys, such as the Two Degree Field Galaxy Redshift Survey (2dFGRS, \citealt{Colless_2001, Cole_2005}), the Galaxy and Mass Assembly (GAMA, \citealt{Driver_2011}), the VIMOS Public Extragalactic Redshift Survey (VIPERS, \citealt{Torre_2013}), the Baryon Oscillation Spectroscopic Survey (BOSS, \citealt{Dawson_BOSS}), and the Extended Baryon Oscillation Spectroscopic Survey (eBOSS, \citealt{eBOSS_Dawson}), have provided large galaxy data that trace large-scale structure and enable stringent constraints on cosmological parameters. In the near future, current and upcoming cosmological surveys will provide even more extensive data with higher accuracy and precision, offering insights into the analysis of large-scale structure of the universe on an unprecedented scale, including the Dark Energy Spectroscopic Instrument (DESI; \citealt{DESI_2016}), the Subaru Prime Focus Spectrograph (PFS; \citealt{Takada_2014}),  the Vera C. Rubin Observatory's Legacy Survey of Space and Time, the Euclid mission (\citealt{Laureijs_2011, Laureijs_2012}) from the European Space Agency (ESA), and the Nancy Grace Roman Space Telescope (\citealt{Green_2012, Dressler_2012, Spergel_2015, Wang_2021}) from NASA.

Cosmological information within large-scale structure can be extracted through various statistics. One of the most fundamental statistics is galaxy clustering, which characterizes the spatial distribution of galaxies. Another key statistic is galaxy--galaxy lensing, which provides insights into the distribution of the underlying dark matter. In the standard $\Lambda$ cold dark matter ($\Lambda$CDM) cosmological model, the parameters $\Omega_m$ and $\sigma_8$ are particularly sensitive to measurements of the large-scale structure. Their combination $S_{8}=\sigma_8(\Omega_m/0.3)^{0.5}$, is effectively constrained by analyses of large-scale structure. Recent studies have reported a discrepancy between large-scale structure measurements at low redshift and observations of the cosmic microwave background (CMB) at high redshift that the CMB measurements generally favor a higher value of $S_{8}$, often referred to as the ``$S_{8}$ tension'' with a significance of $\sim2-3\sigma$ (\citealt{Wibking_2020,Heymans_2021,Abdalla_2022,Abbott_2022,Miyatake_2023,Lange_2023,Hahn_2024,Zhang_2025}). Another crucial cosmological observable, the linear growth rate parameter combination $f\sigma_{8}$, is sensitive to galaxy clustering via the redshift-space distortions (RSD) effect. Similar to $S_{8}$, CMB observations suggest a larger $f\sigma_8$ value compared to several large-scale structure measurements, with a tension of approximately $2\sigma$ (\citealt{Abdalla_2022,Yuan_2022,Lange_2021,Zhai_2023}). On the other hand, we note that several recent studies report closer $S_8$ measurements with CMB after calibration of photometric redshift, indicating that the observed tension may be partially attributed to residual systematics (\citealt{Wright_2025,Janvry_2025}). Therefore the significance of this tension itself is worth further investigation from multiple aspects including the model construction, data analysis and so on.

On large scales, linear perturbation theory provides a simple and reasonably accurate description of the cosmic evolution. However, on smaller scales, non-linear dynamics and complex baryonic processes pose significant challenges to achieving accurate results with simple methods. To extract cosmological information from the non-linear regime, several approaches have been developed, including effective field theory (\citealt{Piazza_2013, Kokron_2022, Cabass_2023}) and simulation-based methods(\citealt{Lange_2021,Lange_2023,Chen_2025}), the latter of which is employed in this work. Using dark matter halos identified in $N$-body simulations, galaxy--halo connection models can be applied to generate mock galaxy catalogs for comparison with observational data. Two widely used empirical approaches in this context are sub-halo abundance matching (SHAM, \citealt{Conroy_2006,Simha_2012,reddick_etal:13,Chaves_2016,Lehmann_2017}) and the halo occupation distribution (HOD, \citealt{HOD_Weinberg,Kravtsov_2004,Zheng_2005,Guo_2016,Yuan_2021}). In the basic HOD framework, galaxy occupation is typically assumed to depend solely on halo mass. However, recent studies suggest that other halo properties or external environment can also influence galaxy occupation (\citealt{Gao_2005,Wechsler_2006,Zentner_2014,Zehavi_2018,Han_2019,Yuan_2020,Xu_2021,Hadzhiyska_2021}), a phenomenon known as galaxy assembly bias or secondary bias. To ensure unbiased cosmological measurements, it is essential to account for this effect when modeling galaxy distribution on non-linear scales.

Combined with galaxy--halo connection models, interpolation methods can be employed to construct emulators that predict various statistics for arbitrary combinations of cosmological and galaxy--halo connection parameters. Emulators are built with a limited number of simulations across the parameter space and have been successfully applied in multiple large-scale structure analyses, utilizing galaxy clustering, galaxy--galaxy lensing and other different summary statistics(\citealt{Wibking_2017,Zhai_2019,Wibking_2020,Yuan_2022,Zhai_2023}). In our earlier work, \cite{Zhai_2019} and \cite{Zhai_2023} (hereafter Z23) developed an emulator approach based on Gaussian Processes (GP) to model galaxy clustering and applied it to analyze the distribution of BOSS galaxies. These studies demonstrated that the emulator approach can provide tight constraints on cosmological parameters through information from the non-linear regime. In this work, we adopt this methodology by incorporating galaxy--galaxy lensing measurements into the analysis. Specifically, we use source galaxies from the Dark Energy Camera Legacy Survey (DECaLS, \citealt{Dey_2019}) and the Hyper Suprime-Cam (HSC, \citealt{Aihara_2022}) to measure galaxy--galaxy lensing signals. This combination enables a more comprehensive exploration of cosmological information from both galaxy clustering and galaxy--galaxy lensing statistics.

In addition to the new summary statistics, we employ the new Aemulus $\nu$ suite (\citealt{DeRose_2023}), a set of $w\nu$CDM $N$-body simulations in this work. The key distinction from earlier Aemulus projects based on previous Aemulus simulations (\citealt{DeRose_2018}) is the inclusion of massive neutrinos as an independent particle species, which can influence the cosmic evolution at different scales. $N$-body simulations provide an ideal framework to explore their distribution and properties (\citealt{Springel_2021, Brandbyge_2009, Ali_2013, Sullivan_2023, Banerjee_2018}). Our model incorporates neutrino mass as a new parameter, allowing us to investigate the impact on large-scale structure of massive neutrinos and potentially constrain its mass through its impact on large-scale structure statistics.

The structure of this paper is as follows: Section \ref{sec:BOSS} introduces the galaxy surveys used in this work. In Section \ref{sec:measurement}, we describe the methods used to measure galaxy clustering and galaxy--galaxy lensing. Section \ref{sec:simulations} details the simulation suites employed in the analysis and the galaxy--halo connection model used to generate galaxy mocks based on these simulations. This section also outlines the likelihood analysis method adopted in this work. Our cosmological constraints and results from various tests are presented in Section \ref{sec:Results}. We provide our conclusion and discussion in Section \ref{sec:conclusion}.

\section{Observational Data} \label{sec:BOSS}

In this work, we use BOSS galaxies for clustering measurements. These galaxies also serve as lens galaxies for galaxy--galaxy lensing analyses, while the source galaxies are selected from the DECaLS and HSC galaxy samples. In this section, we provide a brief introduction to these three galaxy samples.

\subsection{BOSS}

Following \cite{Alam_2017}, we use a combination of the LOWZ and CMASS samples to cover a broad redshift range. The sample is the same as Z23, including the redshift cuts and sample selection. We refer the readers to this paper for more details and we only provide a brief description here.

We divide the galaxy sample into three redshift bins: $0.18<z<0.32$ (low-$z$), $0.32<z<0.48$ (mid-$z$) and $0.48<z<0.62$ (high-$z$), and analyze each bin separately. Within each redshift bin, galaxies are re-selected to maintain a constant number density based on their $i$-band brightness, resulting in galaxy samples that are approximately volume-limited. The final number densities are $2.5\times10^{-4}[h^{-1}\text{Mpc}]^{-3}$ for the low-$z$ bin, and $2.0\times10^{-4}[h^{-1}\text{Mpc}]^{-3}$ for both the mid-$z$ bin and the high-$z$ bin. For clustering measurements, we apply the method from \cite{Guo_2012} to correct for fiber collision effects, where two fibers cannot be placed within an angular separation of 62". For galaxy--galaxy lensing measurements, we use the nearest-neighbor method for corrections (\citealt{Miyatake_2015}).

\subsection{Lensing data}
To measure the galaxy--galaxy lensing signal, we utilize shear catalogs derived from the Fourier\_Quad (FQ) pipeline (\citealt{Zhang_2015}), which incorporates data from the DECaLS (\citealt{Zhang_2022}) and the third public data release of the HSC (\citealt{Liu_2024}). The FQ shear catalogs contain essential information about background galaxies, including their positions, photometric redshifts 
(photo-$z$), magnitudes (mag), signal-to-noise ratios $\nu_F$ (defined in \citet{Li_2021}), and shear estimators. To minimize contamination between background and foreground galaxies due to uncertainties in photo-$z$, we select source galaxies with redshifts $z_s > z_l + 0.2$, where $z_s$ and $z_l$ are the redshifts of the source and lens galaxies, respectively. The FQ shear estimators are derived from the multipole components of the galaxy power spectrum and consist of five distinct estimators: \( G_1 \), \( G_2 \), \( N \), \( U \), and \( V \). Here, $G_i$ are analogous to the ellipticity components $e_i$, $N$ acts as a normalization factor, and $U$ and $V$ serve as additional correction terms, as detailed in \cite{Zhang_2017}. 

Additionally, since \citet{2024arXiv240617991S} find that shear bias varies with redshift, we also conduct an onsite shear bias test for the background galaxies in each galaxy--galaxy measurement with different redshift, known as the field distortion test. This method, proposed by \citet{Zhang2019ApJ}, uses the intrinsic distortions of the CCD focal plane (field distortions) to detect multiplicative and additive biases in the measurements. Finally, we find that the multiplicative biases for our background galaxy sample are small and consistent with zero within $1\sigma$. The additive biases could be removed  by subtracting the signal from random points, where the number of random points is 10 times the number of lenses. Thus, no further correction for shear is needed.

\subsubsection{DECaLS}
DECaLS is one of the surveys within the DESI Legacy Imaging Surveys and provides optical imaging of the DESI footprint in the $g,r$, and $z$ bands, covering approximately $\sim 10,000$ deg$^2$. The photo-$z$ are obtained from \citet{Rongpu2021MNRAS}. For this work, we use the $z$-band shear catalog, which has a galaxy number density of  $\sim3$--$5$ per arcmin$^{2}$, and is shown to have the best imaging quality in \cite{Zhang_2022}. Galaxies are selected with  $\nu_F >4$ and ${\rm mag} < 21$ measurement accuracy, as described in \cite{Liu_2023}. In total, around $9.7\times 10^7$ galaxy images are selected.

\subsubsection{HSC}

HSC provides galaxy images in the $g,r,i,z$ and $y$ bands with the highest quality observed in the $i$-band, covering approximately $\sim 1400$ deg$^2$. Following \cite{Liu_2024}, we select the $r$, $i$, and $z$-band samples to enhance the signal-to-noise ratio in the lensing analysis. We also apply a threshold of $\nu_F >4$ to improve the accuracy of the measurements. The photo-$z$ of HSC galaxies are determined using the DEmP method (\citealt{Nishizawa_2020}), with an associated uncertainty $\sigma_z$. Due to the deep observations of HSC, the highest redshifts of background galaxies can reach approximately $z \sim 5$. However, the photo-$z$ uncertainties for these high-redshift galaxies can be large, potentially introducing biases into the lensing measurements. To address this, we use HSC galaxies in two distinct ways: In the first set (hereafter referred to as HSC), we include all galaxies regardless of their $\sigma_z$. In the second set (hereafter referred to as HSCz), we restrict the sample to galaxies with $\sigma_z < 0.05$.

\section{Measurement of summary statistics} \label{sec:measurement}

In this work, we quantify the clustering measurements of both BOSS galaxies and simulated data sets using the two-point correlation function (2PCF), which includes the projected correlation function $w_{p}$, the redshift-space monopole 
$\xi_{0}$, and the quadrupole $\xi_{2}$. For the galaxy--galaxy lensing measurement, we compute the excess surface density $\Delta \Sigma$. We introduce the measurement of such statistics in this section. 

\subsection{Galaxy clustering}

The 2PCF $\xi(r)$ measures the excess probability of finding galaxy pairs with separation $r$. In practice, the separation is split into two components: along the line of sight ($\pi$) and perpendicular to it ($r_p$). To mitigate the RSD effect caused by peculiar velocities, we use the projected correlation function, defined as
\begin{equation}
w_{p}(r_{p})=2\int_{0}^{\infty}d\pi\xi(r_{p}, \pi).
\end{equation}
We truncate the integral at $\pi_{\rm max}=80h^{-1}$Mpc , following our earlier work, and it has been shown that this choice provides stable results. 

In addition to the projected correlation function, we can alternatively expand the two-point correlation function using Legendre polynomials to obtain the multipoles of different orders: 
\begin{equation}
\xi_{\ell}(s) =\frac{2\ell+1}{2}\int_{-1}^{1} L_{\ell}(\mu)\xi(s, \mu)d\mu,
\end{equation}
where $s$ corresponds to separation in redshift space and $\mu=r_p/s$. In this work, we adopt the monopole $\xi_0$ and quadrupole $\xi_2$ to analyze the cosmological information behind the clustering in redshift space. 

We measure the 2PCF of BOSS galaxies using the estimator from \cite{LS_1993}:
\begin{equation}\label{eq:LS}
\xi(r_{p}, \pi)=\frac{DD-2DR+RR}{RR}.
\end{equation}
Here $DD$, $DR$, and $RR$ the normalized pair counts of data-data, data-random, and random-random pairs, respectively, at the corresponding separations. We use the Planck 2015 cosmology (\citealt{Planck_2015}) to convert redshifts to distances. 
With the 2D measurement, we decompose the signals into $w_{p}$ and $\xi_{\ell}$  over the range 0.1 to 60.2 $h^{-1}$Mpc with 9 logarithmically spaced bins.

\subsection{Galaxy--galaxy lensing}
Galaxy--galaxy lensing is the correlation between the positions of objects, such as galaxies or galaxy clusters, and the surrounding shear, which can be used to measure the Excess Surface Density (ESD) around these objects. For a circularly symmetric lensing potential, the relationship between the ESD and the tangential shear \(\gamma_t\) is given by
\begin{equation}
\Delta \Sigma (R)=\overline{\Sigma} (<R)-\Sigma(R)=\Sigma_c \gamma_t(R).
\end{equation}
Here $\overline{\Sigma}(<r)$ refers to the average surface density within a radius $r$, and $\Sigma_c$ is the comoving critical surface density, defined as 
\begin{equation}\label{sigmac}
\Sigma_c=c^2 D_{\rm s}/[4 \pi G D_{\rm l} D_{\rm l s}(1+z_{\rm l})^2],
\end{equation} 
where $c$ is the speed of light, $G$ is the gravitational constant, $z_{\rm l}$ is the lens redshift, and $D_{\rm s}, D_{\rm l}$, and $D_{\rm ls}$ are the angular diameter distances for the source, lens, and lens-source systems, respectively. Therefore, we can measure the ESD by stacking the tangential shear signals around galaxies. 

In our work, we employ the PDF-symmetrization (PDF-SYM, \citealt{Zhang_2017}) method. The core idea is to construct the Probability Distribution Function (PDF) of the shear signal and determine the corresponding optimal shear value by adjusting the shear values to maximize the symmetry of the PDF. This method maximizes the use of statistical information from the shear estimators and helps reduce statistical biases caused by the uneven distribution or finite number of background sources. For the ESD signal at a certain radius \(r\), we can obtain the distribution \(P(G_{t/\times})\) of all shear estimates \(G_{t/\times}\) for the background galaxies around each lensing galaxy.  However, this distribution deviates from symmetry due to the shear signals. To correct for this, we adjust \(P(G_t)\) by guessing the ESD signal \(\widehat{\Delta \Sigma}\):
\begin{equation}\label{Ghat}
\hat{G}_t=G_t-\frac{\widehat{\Delta \Sigma}}{\Sigma_c}\left(N \pm U_i\right).
\end{equation}
When $\widehat{\Delta \Sigma}$ equals the true ESD signal, the new distribution $P(\hat{G}_t)$ will achieve its most symmetric state. We use $\chi^2$ (defined in Eq. 36 of \citealt{Zhang_2017}) to evaluate the symmetry of the PDF. Additionally, we measure the cross-shear component to detect B-mode signals, and we find that the cross-shear is consistent with zero. This indicates that our shear measurements are not affected by significant systematic errors. To be consistent with the clustering measurement in this work, we divide the radial range from 0.1 to 60.2 $h^{-1}$Mpc into 9 bins in logarithmic space for the lensing measurements.

\section{Models} \label{sec:simulations}

In this section, we describe the simulations used to build and test the emulators, the galaxy--halo connection model for small-scale analysis, and the likelihood method employed in the analysis.

\subsection{Simulations}

We use three suites of $N$-body simulations for different purposes: the Aemulus $\nu$ suite to construct and test the emulators, and the Uchuu\footnote{\url{http://www.skiesanduniverses.org/Simulations/Uchuu/}} \citep{Ishiyama_2020} and UNIT\footnote{\url{https://unitsims.ft.uam.es}} \citep{Chuang_2019} simulations to validate our model. Basic information about these suites is provided below, with more detailed descriptions available in the corresponding papers. 

\subsubsection{Aemulus $\nu$}

\begin{deluxetable*}{l l c r}
\tablecaption{Cosmological parameters adopted in our model, their meaning, and the range for each parameter.\label{tab:param}}
\tablehead{
\colhead{} &
\colhead{Parameter} &
\colhead{Meaning} &
\colhead{Range}
}
\startdata
Cosmology & $n_s$        & Spectral index of the primordial power spectrum                & [0.931, 1.008] \\
      & $H_0$        & Hubble constant                                                & [59.97, 74.27] \\
      & $w$          & Dark energy equation of state                                  & [$-$1.27, $-$0.73] \\
      & $\omega_b$   & Baryon energy density                                          & [0.020, 0.025] \\
      & $\omega_m$   & Matter energy density                                          & [0.110, 0.130] \\
      & $M_{\nu}$    & Total neutrino mass [eV]                                           & [0.011, 0.442] \\
      & $\sigma_8$   & Matter fluctuations on 8 $h^{-1}$ Mpc predicted by linear theory            & [0.680, 0.940] \\
\enddata
\end{deluxetable*}

Similar to previous Aemulus simulations, Aemulus $\nu$ is a new suite of $N$-body simulations that includes massive neutrinos as an independent particle species. Neutrino mass is treated as a new parameter when constructing the emulators, allowing us to obtain constraints from observational data. There is a two-tiered parameter space design for Aemulus $\nu$ simulations. Tier 1 includes 100 simulations across a broader parameter space, while Tier 2 consists of 50 simulations with a tighter parameter distribution. To achieve higher accuracy in our predictions, we use the Tier 2 simulations to construct the emulators, which provides coverage similar to that of the Aemulus simulations in the parameter space. Each Aemulus $\nu$ simulation contains $1400^3$ dark matter particles and $1400^3$ neutrino particles within a box of side length $1050~h^{-1}$Mpc. The resulting mass resolution is sufficient for modeling massive galaxies, as required in this work.

The simulation suite adopts a $w\nu$CDM cosmology with parameters including the scalar spectral index $n_s$, the Hubble parameter $H_0$, the dark energy equation of state parameter $w$, the dark matter density $\omega_c$, the baryon density $\omega_b$, the amplitude of matter fluctuations $\sigma_{8}$ and the total neutrino mass $M_{\nu}$ . Table \ref{tab:param} provides a summary of these parameters including their ranges in our model.  Similar to Aemulus V (Z23), we also introduce $\gamma_f$ as an additional parameter to scale the amplitude of the halo velocity field relative to GR (\citealt{Reid_2014}), thus enabling a constraint on the RSD effect. 
Since no specifically designed test simulations are available, as in the previous Aemulus simulations, we choose sim 1-40 from Tier 2 simulations of Aemulus $\nu$ to train our emulators, reserving the remaining 10 simulations for testing their accuracy. We employ simulations at redshifts $z=0.2700,0.3974,0.5375$ to model the observational data in three redshift bins respectively.

\subsubsection{Uchuu and UNIT}

We adopt the HOD model and galaxy assembly bias based on environment to generate galaxy mocks and construct the emulators. To evaluate the robustness of our model, we create galaxy mocks using the Uchuu and UNIT simulations with the SHAM model. We then perform recovery tests using our emulators, similar to the analysis in Aemulus V, but for both galaxy clustering and galaxy--galaxy lensing. The SHAM model assumes that galaxies reside in halos or subhalos, with a correlation between the stellar mass or luminosity of a galaxy and the mass or velocity of its host dark matter halo or subhalo (\citealt{Conroy_2006}). Due to the mapping process, the SHAM model can introduce various amount of galaxy assembly bias. Therefore, recovery tests of our HOD-based model on these SHAM galaxy mocks allow us to examine potential systematics in cosmological constraints arising from different galaxy--halo connection models.

Both the Uchuu and UNIT simulations adopt the Planck 2015 cosmology. The Uchuu simulation contains $12800^3$ dark matter particles with a mass of $3.27\times10^{8}h^{-1}M_{\odot}$. The box size of the Uchuu simulation is $2000 ~h^{-1}$Mpc, around 8 times the volume of the Aemulus $\nu$ simulations. The UNIT simulations employ the inverse phase technique (\citealt{Angulo_2016}) to reduce cosmic variance. For our analysis, we use two pairs of UNIT simulations, each with $4096^3$ dark matter particles in a $1~h^{-1}$Gpc box. When generating SHAM mocks, we apply the model of \cite{Lehmann_2017}, which combines the maximum circular velocity and the virial velocity of halos as the quantity on which to rank the subhalos in the abundance matching process. More details are provided in Appendix \ref{appsec:recovery} and Z23. 

When building our emulators, Gaussian Processes require the input error for our training statistics. Since the Aemulus $\nu$ simulations have the same box size as the previous Aemulus simulations, we use the training error of galaxy clustering from Z23 which is estimated from multiple boxes with different initial conditions. For galaxy--galaxy lensing, however, we need particle-level information to calculate $\Delta \Sigma$ which is not directly accessible from the Aemulus simulations. To address this, we divide the Uchuu simulation into 8 subvolumes, each with a size similar to the Aemulus $\nu$ simulations. These subvolumes are then used to evaluate the training error for galaxy--galaxy lensing, which is incorporated into our emulators. The training error and the performance of our emulators are discussed in Appendix \ref{appsec:performance}. For example, the accuracy of $w_p$ at 1 to 10 $h^{-1}$Mpc is at 1-2\% level. The other clustering statistics are slightly worse but comparable to the sample variance. On the other hand, the lensing signal $\Delta \Sigma$ is better and the accuracy can be up to 1\% at a few $h^{-1}$Mpc scales.

\subsection{Galaxy--halo connection} \label{sec:HOD}

In this work, we use the HOD model to characterize the galaxy distribution at non-linear scales and construct galaxy mocks from $N$-body simulations, as in Z23. Following \cite{Zheng_2005}, galaxies are divided into central and satellite populations, each with distinct occupation models. Four parameters, $\log{M_\text{sat}}$, $\alpha$, $\log{M_\text{cut}}$, $M_\text{min}$ and $\sigma_{\log{M}}$, are related to the mean occupation numbers of central and satellite galaxies. Additionally, we introduce the parameter $\eta_\text{con}$ to describe the relationship between the concentration of satellite galaxies and the host dark matter halo. Two more parameters, $\eta_\text{vc}$ and $\eta_\text{vs}$, account for the velocity biases of central and satellite galaxies, respectively. The final parameter, $f_\text{max}$, is used to scale the amplitude of the central galaxy occupation numbers at the massive end.

To model galaxy assembly bias, we introduce an additional set of parameters to capture the effects of the halo environment (\citealt{McEwen_2016}). Specifically, we calculate the relative density $\delta$ within a $10~h^{-1} \text{Mpc}$ radius around each host halo using a Top-Hat model. The HOD parameter $M_{\text{min}}$ is then modified as a function of $\delta$ by the following expression:
\begin{equation}\label{eq:Mmin_AB}
M_{\rm min}(\delta) = M_{\text{min}}\left[1 + f_{\rm env}\text{erf}\left(\frac{\delta-\delta_{\rm env}}{\sigma_{\rm env}}\right)\right],
\end{equation}
where $f_{\rm env}$, $\delta_{\rm env}$, and $\sigma_{\rm env}$ are three parameters that describe the galaxy assembly bias. If we set $f_{\rm env} = 0$, $M_{\text{min}}$ becomes independent of the halo environment, and the model reduces to the basic HOD framework without any galaxy assembly bias.

\subsection{Likelihood analysis} \label{sec:likelihood}

We use simulation-based emulators to make accurate predictions of the summary statistics for a given set of parameters, for both the cosmological and galaxy--halo connection models. By comparing these predictions with observational data, we can place constraints on the model parameters through a likelihood analysis.

In the earlier work, Z23 tested different methods for constructing the covariance matrix, including jackknife resampling, GLAM\footnote{\url{http://www.skiesanduniverses.org/Simulations/GLAM/}} simulations (\citealt{Klypin_2018}) and their combinations, all of which produce consistent results. In this work, we adopt the covariance matrix constructed using the jackknife resampling method for both clustering and lensing. Following \cite{Lange_2023}, we ignore the cross-covariance between clustering and lensing, as it is expected to be negligible (\citealt{Taylor_2022}).

In addition to the sample variance described above, we also account for the contribution from emulator inaccuracies. Following the method used in Z23 for galaxy clustering, we estimate the contribution for galaxy--galaxy lensing by comparing emulator predictions with test simulations. The final covariance matrix used in the likelihood calculation combines these two components. The likelihood function is then computed as

\begin{equation}\label{eq:likelihood}
\ln{\mathcal{L}} = -\frac{1}{2}(\xi_{\text{emu}}-\xi_{\text{obs}})C^{-1}(\xi_{\text{emu}}-\xi_{\text{obs}}),
\end{equation}
where $\xi_{\text{emu}}$ is prediction given by emulators and 
$\xi_{\text{obs}}$ is the observational data, both of which include galaxy clustering and galaxy--galaxy lensing. We perform likelihood analysis on galaxy clustering from 0.1 to 60.2 $h^{-1}$Mpc and galaxy--galaxy lensing from 1.7 to 60.2 $h^{-1}$Mpc in our fiducial analysis. The first 4 lensing measurements at small scales are excluded due to complicated baryonic effect, following the early work (\citealt{Lange_2023}).

In the measurement of clustering and lensing from HOD mocks, the mismatch of mock cosmology and the fiducial cosmology adopted in observational measurement can introduce systematic differences in the inferred signals, known as Alcock-Paczynski (AP) effect (\citealt{Alcock_1979}). However, as shown by \cite{Chapman_2021}, \cite{Lange_2021}, and \cite{Zhai_2024}, this effect on small-scale clustering analysis is insignificant. We therefore neglect an AP correction for clustering. For lensing, we conduct additional tests, and find nearly identical results with and without the correction. More details can be found in Appendix \ref{appsec:ap}. Hence, we do not apply the AP correction at the model level.
 
We use the MultiNest algorithm\footnote{\url{https://github.com/JohannesBuchner/MultiNest}} to perform Bayesian inference. 
As mentioned in \cite{Lemos_2023}, hyperparameter setting is essential for MultiNest. Following Z23, Our nested-sampling setup uses 1000 live points with efficiency=0.8, tolerance=0.5 and constant efficiency mode turned off. We also test other setups for efficiency and tolerance values and find that the result is quite stable. The output provides the posterior distribution as a byproduct. Since the Aemulus $\nu$ simulation has a slightly different cosmological parameter space design compared to Aemulus, we update the multidimensional ellipsoid prior space of cosmological parameters restricted by the distributions of cosmological parameters in the Aemulus $\nu$ Tier 2 simulation suite using the same method outlined in Appendix D of Z23. The prior for $\gamma_f$ and HOD parameters remains flat with a uniform distribution and their ranges are same as in Z23.

We test our HOD-based model on SHAM galaxy mocks generated from Uchuu and UNIT simulations. A likelihood analysis demonstrates that the main cosmological parameters can be well estimated and recovered within $1\sigma$, as detailed in Appendix \ref{appsec:recovery}. Finally, we apply our emulators and likelihood analysis to the BOSS, DECaLS, and HSC data.

\section{Results} \label{sec:Results}

In this section, we present the constraints on the model parameters with various priors, focusing on key measurements such as $f\sigma_8$, $S_{8}$, and neutrino mass. We also compare our results with those from other analyses in the literature that employ various models to extract cosmological information at non-linear scales.

\subsection{Summary statistics fit results}

This work analyzes four separate sets of observational data. The first set considers only galaxy clustering, similar to the analysis in Z23, with the primary difference being the simulation suite used to construct the emulators. The remaining three sets combine galaxy clustering and galaxy--galaxy lensing, with measurements taken using different source galaxy samples.
\begin{figure*}[htbp]
\begin{center}
\includegraphics[width=18.0cm]{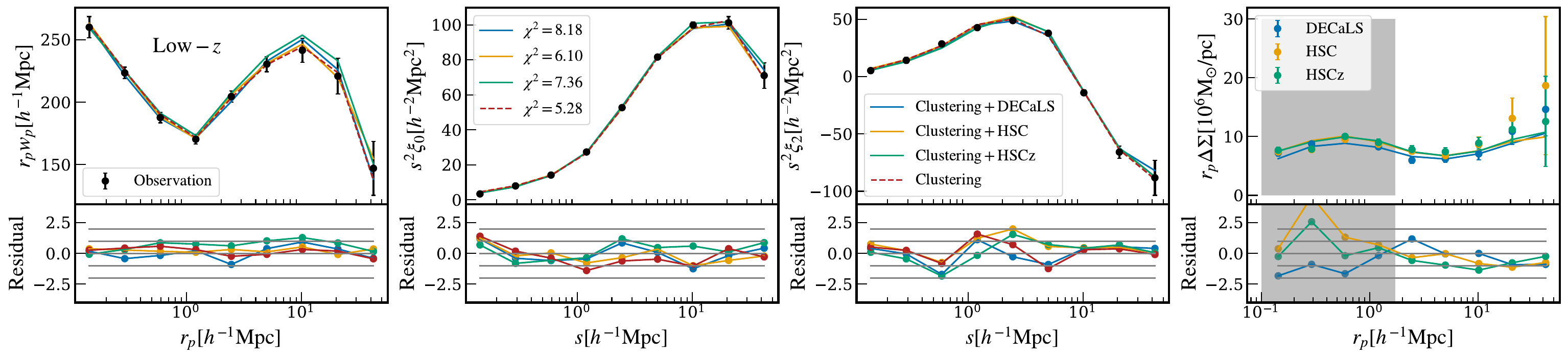}
\includegraphics[width=18.0cm]{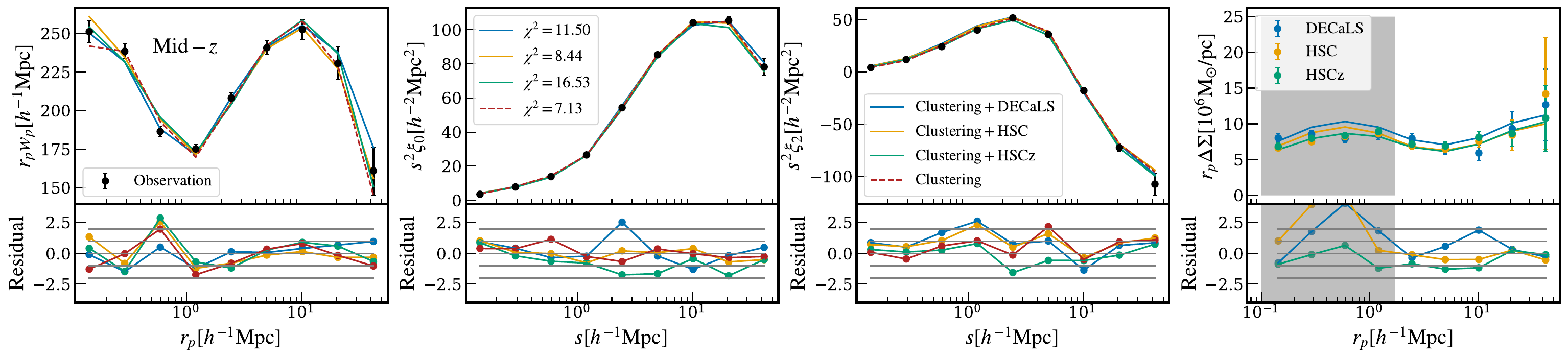}
\includegraphics[width=18.0cm]{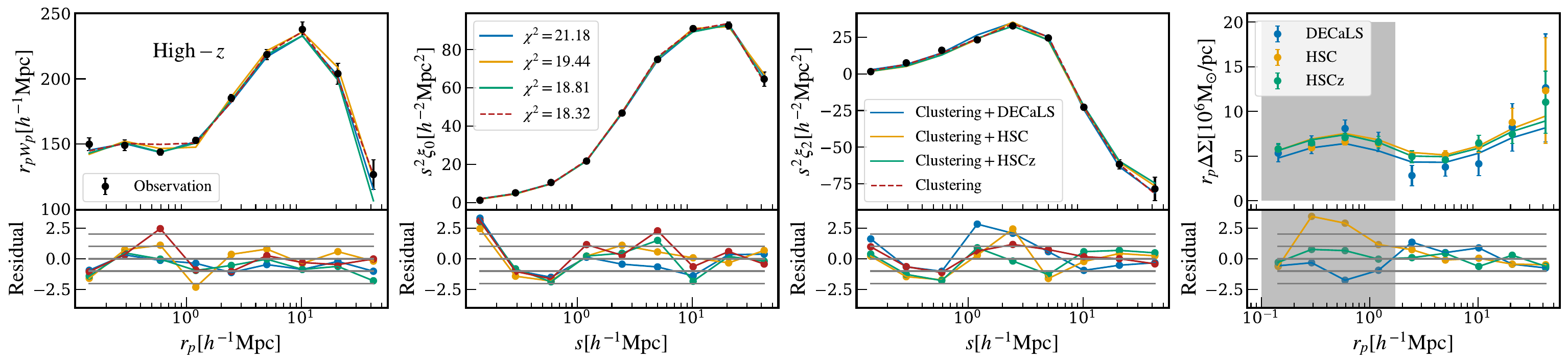}
\caption{Best-fit model of $w_{p}$, $\xi_{0}$, $\xi_{2}$, and $\Delta \Sigma$ at three redshift bins, the total degrees of freedom in the analysis is 13. For the Clustering-only case with dashed red lines, no $\Delta \Sigma$ data are shown. For the Clustering+Lensing sets measured with different source galaxies, the best fits for clustering and lensing are displayed in three colors. The clustering statistics $w_{p}$, $\xi_{0}$, and $\xi_{2}$ are the same across all four combinations of observational data and are shown as black dots. $\Delta \Sigma$ measured with different source galaxies is presented as dots in the corresponding colors. The bottom panel of each figure shows the residuals with respect to the BOSS measurements, normalized by the observed uncertainty, with the horizontal lines indicating the 1$\sigma$ and 2$\sigma$ levels. Lensing data in the grey shaded region corresponds to scales excluded from our fiducial analysis.}
\label{fig:bestfit_BOSS_lowz}
\end{center}
\end{figure*}

\begin{figure*}[htbp]
\begin{center}
\includegraphics[width=8.0cm]{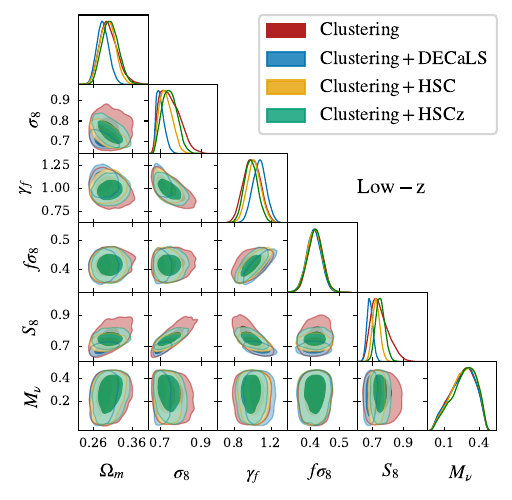}
\includegraphics[width=8.0cm]{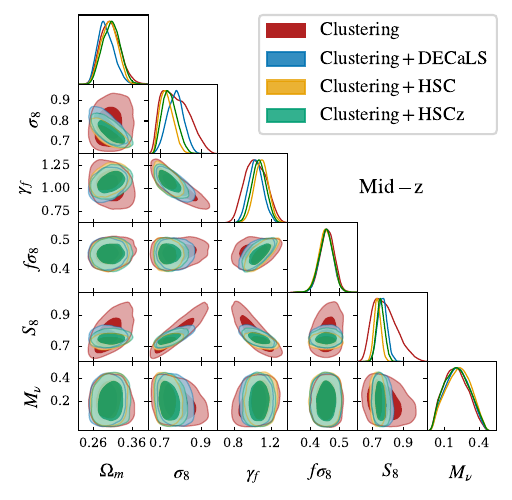}
\includegraphics[width=8.0cm]{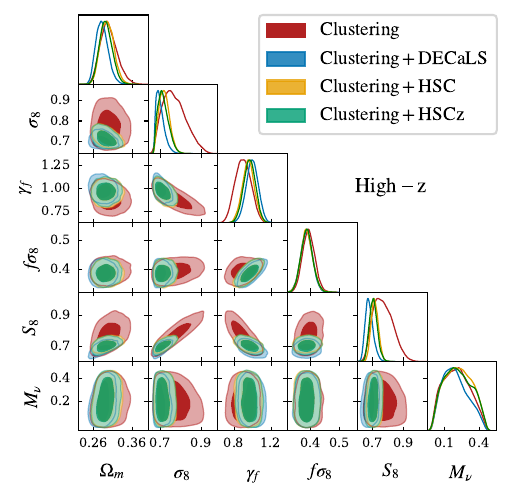}
\caption{Fiducial constraints on a subset of cosmological parameters at different redshifts. The contours represent the 1- and 2-$\sigma$ confidence levels. The results show a comparison of different observational data: Clustering (red) refers to constraint from galaxy clustering data ($w_{p} + \xi_{0} + \xi_{2}$) from BOSS galaxies; Clustering+DECaLS, Clustering+HSC, and Clustering+HSCz represent joint constraints from galaxy clustering data and galaxy--galaxy lensing data ($\Delta \Sigma$) measured with different source galaxies.}
\label{fig:fiducial_constraint}
\end{center}
\end{figure*}
In the fiducial analysis, all parameters are allowed to vary within their priors. 
Figure \ref{fig:bestfit_BOSS_lowz} shows the best-fit results and the minimum $\chi^2$ for each test in different redshift bins. There are 8 cosmological parameters, 8 HOD parameters, 3 galaxy assembly bias parameters, 9 data points for each clustering statistic, and 5 data points for lensing in our fiducial analysis. Therefore, the number of degrees of freedom is 8 for clustering only and 13 for the combination of clustering and lensing. In most cases, we obtain reasonable $\chi^2$. Similar to Z23, we find that the value of $\chi^2$ at high redshift is relatively high. As shown in Figure \ref{fig:bestfit_BOSS_lowz}, this is mainly due to the monopole data at the smallest scale, which accounts for around 10 in the $\chi^2$ calculation. The relationship between $\chi^2$ and degrees of freedom suggests that our analysis provides reasonable results. In Figure \ref{fig:fiducial_constraint}, we present constraints on a subset of cosmological parameters from our fiducial analysis. We do not observe significant tensions between different data sets. The addition of galaxy--galaxy lensing strengthens the constraining power on some parameters, particularly $\sigma_8$ and $S_{8}$, as expected. Since the constraint on $f\sigma_8$ mainly comes from clustering in redshift space, the addition of lensing data narrows the range of $\sigma_8$ and $\gamma_f$, the addition of lensing data narrows the range of $\sigma_8$ and $\gamma_f$, but the compression is along the direction of degeneracy. Hence, the improvement in the $f\sigma_8$ constraint is minor. For instance at high redshift, adding lensing signals in the analysis can improve the constraint from 5.36\% to 5.18\%. In addition, we do not find strong dependence of the results on redshift. Note that the low-redshift sample has somewhat higher density than the other two, which can make the comparison slightly non-trivial.

\begin{figure}[htbp]
\begin{center}
\includegraphics[width=8.0cm]{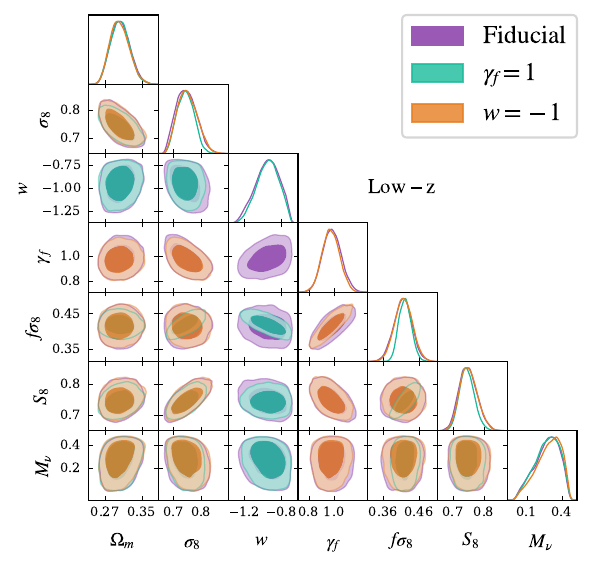}
\caption{Constraints with different priors at low-$z$ using the combination of clustering and lensing measured with HSCz source galaxies. The fiducial results allow all parameters to vary within the prior of our emulators. Fixing $\gamma_{f}=1$ corresponds to General Relativity (GR), while fixing $w=-1$ corresponds to the $\Lambda$CDM cosmology.}
\label{fig:df_prior_constraint}
\end{center}
\end{figure}

In addition to the fiducial analysis, we also examine the constraints on other cosmological parameters by fixing  $w=-1$ or $\gamma_f=1$. Figure \ref{fig:df_prior_constraint} illustrates the constraints obtained from galaxy clustering and galaxy--galaxy lensing measured with HSCz at low redshift, under these different priors. We find that the constraints on the key cosmological parameters are consistent with the fiducial analysis, with no significant offsets greater than $1\sigma$. The results for the other redshift bins and data combinations show similar consistency, aligning well with the previous clustering-only results (Z23). Similar to Z23, we test the cosmological constraints from different scales of clustering and lensing in Appendix \ref{appsec:scale}. Our results from different combinations of observational data prefer galaxy assembly bias parameter $f_{\rm env}=0$. We further investigate the effect of galaxy assembly bias in Appendix \ref{appsec:bias}.

\subsection{Lensing systematics}

In addition to fiber collisions, galaxy lensing measurements are also subject to systematic effects arising from photo-$z$ shift, the potential systematics from shear measurements, and others (\citealt{Miyatake_2023}).

These systematic effects can be mitigated by multiplying $\Delta \Sigma$ by a correlation factor that accounts for the contributions of different sources. For simplicity, we employ an additional parameter to approximately correct for these effects. During the likelihood analysis, we uniformly scale the $\Delta \Sigma$ predicted by our emulators by a factor $A_\text{lens}$, with a uniform prior, following \cite{Miyatake_2023}. The prior range for $A_\text{lens}$ is set to 0.2 -- 2.0. Tests with narrower priors showed that the posterior tended to hit the boundaries, indicating that tighter ranges would impose informative constraints and potentially bias the inference. We therefore adopt a broad prior to ensure that the amplitude is primarily driven by the data.

\begin{figure}[htbp]
\begin{center}
\includegraphics[width=8.0cm]{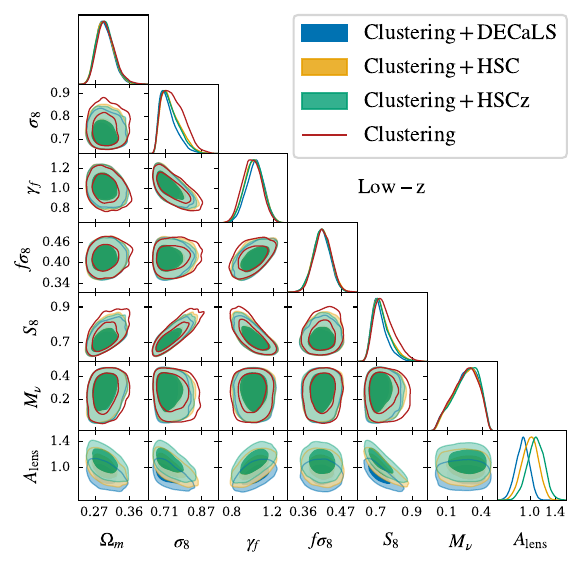}
\caption{Cosmological constraints with $A_\text{lens}$ as a free parameter at low-$z$.  As in the fiducial results, different colors represent different observational data sets. The $A_\text{lens}$ parameter only influences the lensing data and therefore has no effect on the clustering-only results; the red contours for clustering-only are identical to those in the fiducial analysis and serve as a reference. For lensing data measured with different source galaxies, the final row displays the posterior distributions of $A_\text{lens}$.}
\label{fig:delta_constraint}
\end{center}
\end{figure}

In Figure \ref{fig:delta_constraint}, we present cosmological constraints along with the posterior distributions of $A_\text{lens}$ for different observational data sets at low redshift. Since $A_\text{lens}$ only affects galaxy--galaxy lensing, the red contours representing the clustering-only result are the same as those in the fiducial analysis and serve as a reference. We can see that the constraint on $A_\text{lens}$ varies by $\sim1-2\sigma$ depending on the lensing data set. Since $S_{8}$ is more sensitive to lensing data, it is degenerate with this multiplicative factor, while minimal impacts are seen for other key cosmological parameters. We also see similar behaviors at mid-$z$ and high-$z$ when considering the $A_\text{lens}$ parameter. For clarity, we do not present the full set of cosmological constraints for these redshifts here, but instead focus on its impact on $S_{8}$, as detailed in Section \ref{sec:S8}. 

Meanwhile, the wide prior of $A_\text{lens}$ may introduce the prior-volume effect (\citealt{Guachalla_2025}), which can shift the posterior distribution of correlated parameters, such as $S_{8}$, and potentially bias their inferred values. In order to examine the possible prior-volume effect, we conduct an additional recovery test with $A_\text{lens}$ marginalized over on the Uchuu SHAM mock, and find that the impact from prior-volume effect is not significant. We provide more details in Appendix \ref{appsec:recovery}.

\subsection{Constraints on $f\sigma_8$ and $S_{8}$}\label{sec:S8}

Based on the posterior distributions of the parameters, we extract key cosmological measurements from the joint analysis of galaxy clustering and lensing, specifically the linear growth rate parameter combination $f\sigma_{8}$ and the structure amplitude $S_{8}$.

\begin{figure*}[!htbp]
\begin{center}
\includegraphics[width=16.0cm]{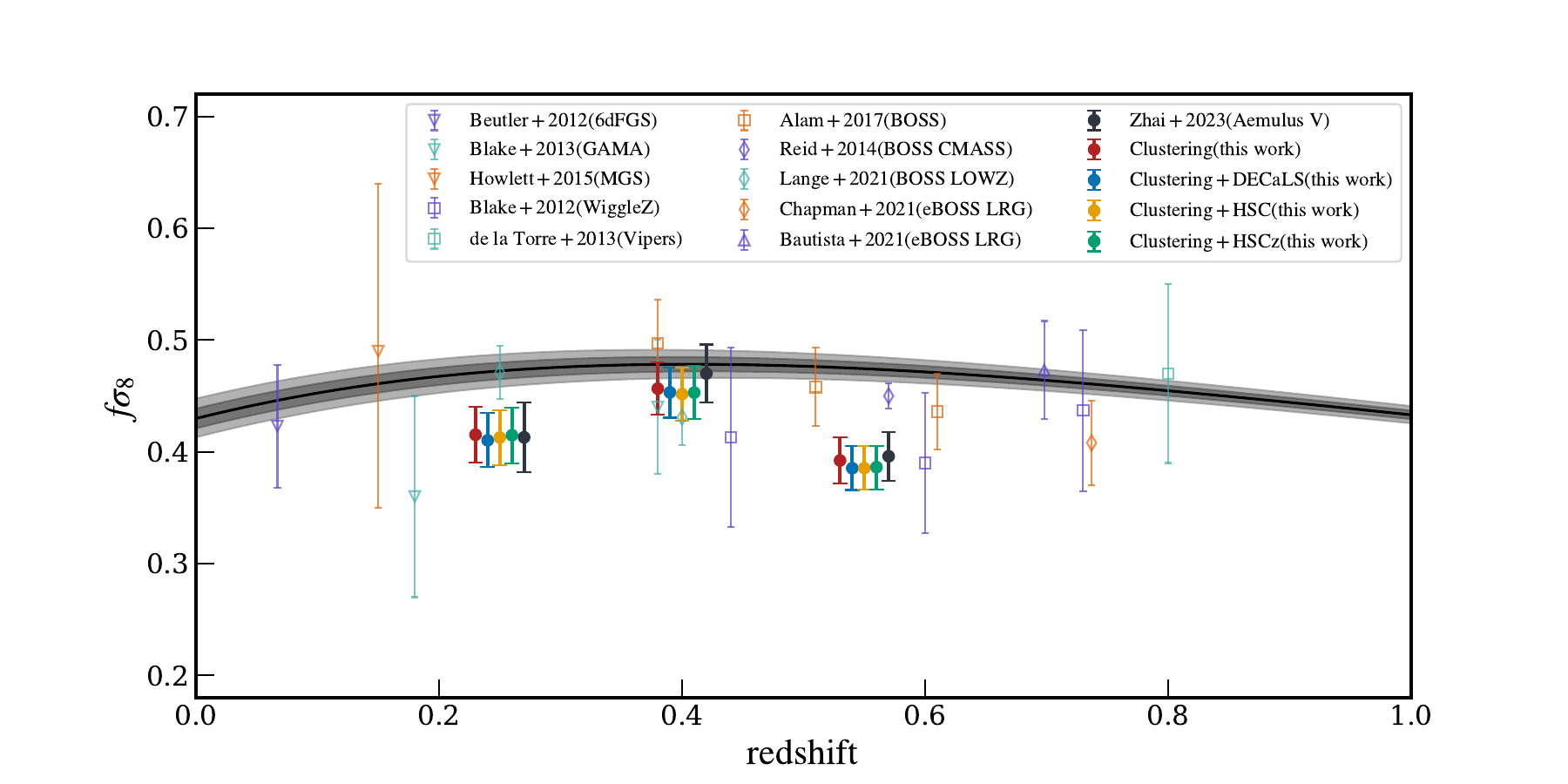}
\caption{Fiducial constraints on $f\sigma_8$ at three redshifts. Black dots represent result from Aemulus V (Z23). Planck 2018 result for $f\sigma_8$ at different redshifts is shown as black line with 1$\sigma$ and 2$\sigma$ uncertainty. We also present some recent results of $f\sigma_8$ from different galaxy surveys, including 6dFGS (\citealt{Beutler_2012}), GAMA (\citealt{Blake_2013}), SDSS-I/II main galaxy sample (\citealt{Howlett_2015}, MGS), WiggleZ (\citealt{Blake_2012}), Vipers (\citealt{de_la_Torre_2013}), SDSS DR12 (\citealt{Alam_2017}), SDSS CMASS (\citealt{Reid_2014}), SDSS LOWZ (\citealt{Lange_2021}) and eBOSS-LRG (\citealt{Bautista_2021,Chapman_2021}). Our measurements at the same redshifts are shifted slightly for plotting purpose.}
\label{fig:fsigma8_constraint}
\end{center}
\end{figure*}

As mentioned earlier, the inclusion of lensing data does not significantly break the degeneracy of $f\sigma_8$, resulting in only a minor improvement in the constraints, as summarized in Table \ref{tab:fsigma8}. To visualize the tension, we present our results alongside those from Z23 and the Planck 2018 measurements (\citealt{Planck_2020}) at different redshifts in Figure \ref{fig:fsigma8_constraint}, as well as several recent results from other galaxy surveys. Compared to Z23, our analysis yields similar values at low redshift, but slightly smaller values at middle and high redshifts. Since the uncertainties of our constraints are smaller, we observe slightly larger tensions with the Planck 2018 results, which are $0.472\pm0.007$, $0.478\pm0.006$ and $0.474\pm0.006$ at low-$z$, mid-$z$ and high-$z$ respectively.

For the $S_{8}$ measurement, galaxy--galaxy lensing is considered a crucial probe (\citealt{Wibking_2020}, \citealt{Lange_2023}). The first part in Table \ref{tab:S8} presents our fiducial constraints on $S_{8}$. Lensing data can significantly reduce the uncertainty. At high redshift, the constraint is improved from 8.3\% to 3.5\% when the HSCz lensing data is added into the analysis. The other lensing data at different redshifts show similar performance. In addition, the results also favor a smaller $S_{8}$ value compared to $0.834\pm0.016$ of Planck 2018 result, consistent with previous findings of the ``$S_{8}$ tension'' between large-scale structure analysis of the late universe and the CMB analysis of the early universe. 

As shown in Table \ref{tab:S8}, we obtain the lowest $S_{8}$ when combining clustering and lensing measured with DECaLS galaxies at high redshift, while considering that the $A_\text{lens}$ parameter increases the estimates of $S_{8}$ from lensing measured with DECaLS to a level consistent with the measurements from the other data sets. In addition, the inclusion of  $A_\text{lens}$ weakens the overall constraint on $S_{8}$ . The final constraint on $S_{8}$, when considering the $A_\text{lens}$ parameter, thus yields a smaller tension with Planck, as presented in the second part of Table \ref{tab:S8}. However, even after accounting for various potential systematic effects, the inclusion of lensing data still favors the persistence of this tension in our analysis. It is also worth mentioning that larger uncertainty of $A_\text{lens}$ will lead to weaker constraints on $S_{8}$ considering their degeneracy, and hence it is essential to achieve more accurate calibration for lensing systematics, especially for the main source photo-$z$ shift, in order to obtain a tighter constraint on $S_{8}$.

\begin{table}[!htbp]
\caption{Constraints on $f\sigma_8$ with 1$\sigma$ uncertainty at three redshifts from different data in the fiducial case. Result from Z23 is shown as Aemulus V in the first row.}
\begin{center}
\begin{tabular}{rccc}
\cline{1-4}
Observational data      & low-$z$ & mid-$z$ & high-$z$  \\
\cline{1-4}
Aemulus V   & $0.413^{+0.031}_{-0.031}$  & $0.470^{+0.026}_{-0.026}$  & $0.396^{+0.022}_{-0.022}$ \\
Clustering   & $0.415^{+0.025}_{-0.025}$  & $0.456^{+0.024}_{-0.024}$  & $0.392^{+0.021}_{-0.020}$ \\
Clustering+DECaLS  & $0.410^{+0.024}_{-0.024}$  & $0.453^{+0.022}_{-0.023}$  & $0.385^{+0.020}_{-0.020}$ \\
Clustering+HSC & $0.413^{+0.025}_{-0.025}$  & $0.452^{+0.024}_{-0.024}$  & $0.385^{+0.020}_{-0.019}$ \\
Clustering+HSCz  & $0.415^{+0.025}_{-0.025}$  & $0.453^{+0.024}_{-0.024}$  & $0.386^{+0.019}_{-0.020}$ \\
\cline{1-4}
\end{tabular}
\end{center}
\label{tab:fsigma8}
\end{table}

\begin{table}[!htbp]
\caption{Constraints on $S_{8}$ with 1$\sigma$ uncertainty at three redshifts from different data. The top section shows the fiducial constraints, and the bottom section shows the results when including the $A_\text{lens}$ parameter.}
\begin{center}
\begin{tabular}{rccc}
\cline{1-4}
Fiducial      & low-$z$ & mid-$z$ & high-$z$  \\
\cline{1-4}
Clustering   & $0.738^{+0.066}_{-0.045}$  & $0.770^{+0.087}_{-0.067}$  & $0.762^{+0.069}_{-0.057}$ \\
Clustering+DECaLS  & $0.680^{+0.019}_{-0.017}$  & $0.761^{+0.030}_{-0.030}$  & $0.675^{+0.027}_{-0.022}$ \\
Clustering+HSC & $0.721^{+0.025}_{-0.023}$ & $0.729^{+0.022}_{-0.021}$  & $0.711^{+0.024}_{-0.022}$  \\
Clustering+HSCz   & $0.747^{+0.030}_{-0.027}$  & $0.749^{+0.024}_{-0.026}$  & $0.703^{+0.025}_{-0.024}$ \\

\cline{1-4}
Considering $A_\text{lens}$      & low-$z$ & mid-$z$ & high-$z$  \\
\cline{1-4}
Clustering+DECaLS  & $0.711^{+0.050}_{-0.032}$  & $0.806^{+0.078}_{-0.077}$  & $0.739^{+0.067}_{-0.050}$ \\
Clustering+HSC & $0.721^{+0.059}_{-0.037}$  & $0.736^{+0.071}_{-0.046}$  & $0.731^{+0.059}_{-0.044}$ \\
Clustering+HSCz   &  $0.719^{+0.056}_{-0.036}$ & $0.740^{+0.078}_{-0.049}$  & $0.738^{+0.063}_{-0.047}$ \\

\cline{1-4}
\end{tabular}
\end{center}
\label{tab:S8}
\end{table}

\begin{figure}[!htbp]
\begin{center}
\includegraphics[width=8.4cm]{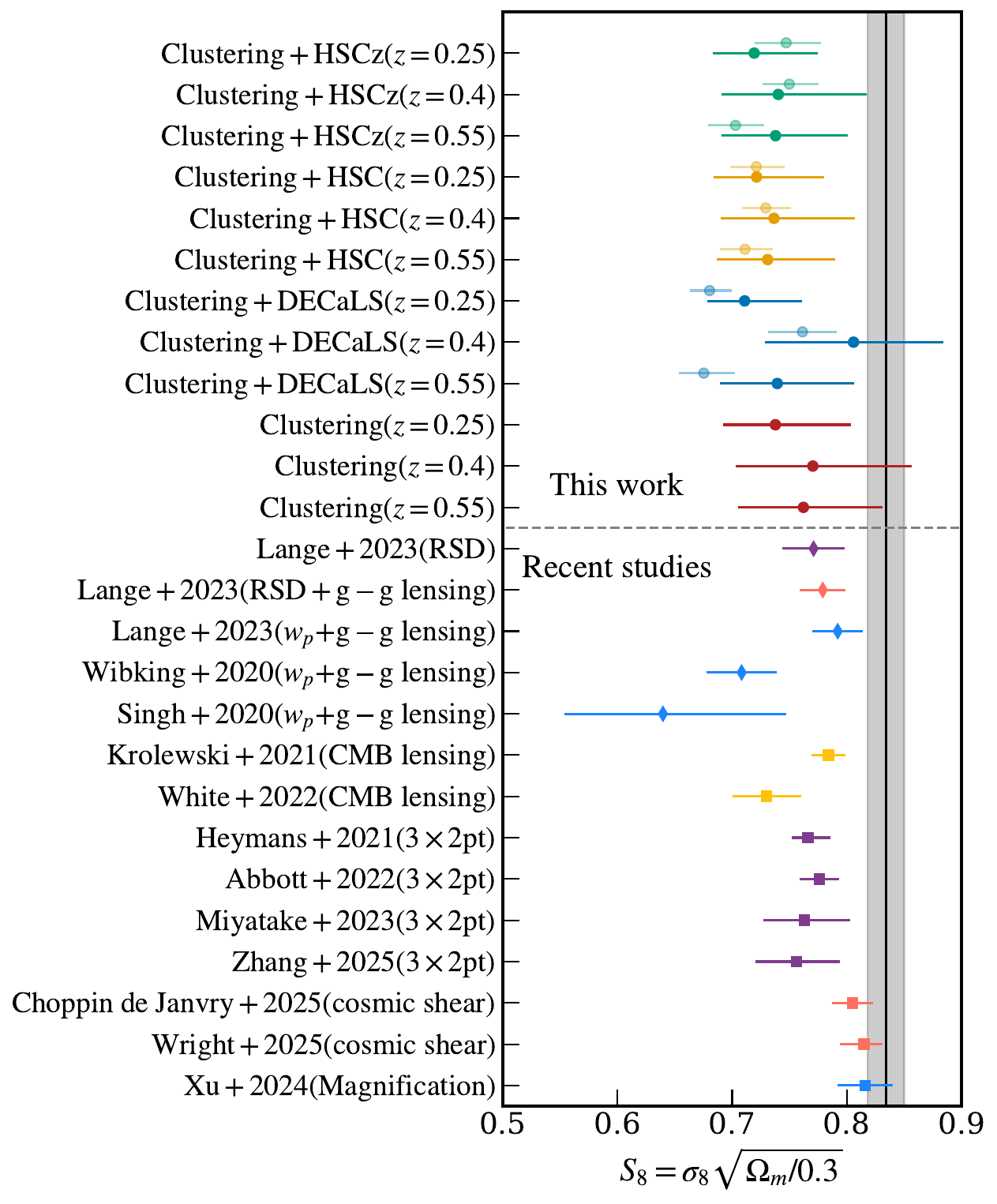}
\caption{Measurement of $S_{8}$ using different data sets at three redshifts. Planck 2018 result is shown as black line with 1$\sigma$ shaded area. The top panel shows results from our work, where dots with higher transparency correspond to fiducial results(i.e., galaxy clustering from 0.1 to 60.2 $h^{-1}$Mpc and galaxy--galaxy lensing from 1.7 to 60.2 $h^{-1}$Mpc), while dots with lower transparency represent constraints considering $A_\text{lens}$ parameter. The bottom panel presents the latest results from large-scale structure research using multiple approaches (\citealt{Wibking_2020}, \citealt{Singh_2020b}, \citealt{Krolewski_2021}, \citealt{Heymans_2021},\citealt{White_2022}, \citealt{Abbott_2022}, \citealt{Miyatake_2023}, \citealt{Lange_2023}, \citealt{Xu_2024}, \citealt{Zhang_2025}, \citealt{Wright_2025}, \citealt{Janvry_2025}).} 
\label{fig:S8_constraint}
\end{center}
\end{figure}

Finally, to compare with recent studies on large-scale structure, we present constraints on $S_{8}$ using different methods alongside our results in Figure \ref{fig:S8_constraint}. Similar to our analysis, \cite{Lange_2021}, \cite{Wibking_2020}, and \cite{Singh_2020b} perform joint analyses combining galaxy clustering and galaxy--galaxy lensing based on BOSS LOWZ galaxies as clustering tracers and lens galaxies with various source galaxy  samples, employing distinct approaches. \cite{Krolewski_2021} and \cite{White_2022} analyze correlations of unWISE galaxies and DESI Luminous Red Galaxies with Planck CMB lensing, respectively. \cite{Heymans_2021}, \cite{Abbott_2022}, \cite{Miyatake_2023} and \cite{Zhang_2025} conduct 3 $\times$ 2pt (cosmic shear, galaxy--galaxy lensing and projected galaxy clustering) analyses in different models based on KiDS, DES and HSC source galaxies combined with multiple galaxy samples as lens galaxies and clustering tracers. Above works analyzing large-scale structure using various statistics present a consistent lower result of $S_{8}$ than CMB analysis, as shown in Figure \ref{fig:S8_constraint}, which indicates the existence of so-called ``$S_{8}$ tension.'' Overall, our results are consistent with those from these large-scale structure analyses, and also exhibit $1\sim 4\sigma$ tension with the Planck results. However, it is worth noting that some analyses using different approaches show different results. For instance, estimations of $S_{8}$ from lensing magnification measurements of BOSS CMASS lens galaxies (\citealt{Xu_2024}) and cosmic shear measurements of KiDS galaxies with updated photometric redshift calibration method(\citealt{Wright_2025}) are in good agreement with CMB, and \cite{Janvry_2025} reanalyzes cosmic shear from HSC and finds much closer $S_{8}$ with CMB after calibration of photometric redshift with clustering redshifts method, as shown in Figure \ref{fig:S8_constraint}. \cite{Shao_2023} finds that considering galaxy bias at linear scale, clustering and lensing measurements of BOSS LOWZ galaxies can lead to a measurement of $S_{8}$ consistent with CMB. \cite{Contreras_2023} and \cite{Mahony_2025} employ extended SHAM model to significantly alleviate the tension of $S_{8}$ measured with galaxy clustering and galaxy--galaxy lensing with CMB. Considering different statistics and methodologies adopted in the large scale structure analyses, the (in)consistencies between experiments requires further and more thorough examinations. 

\subsection{Neutrino mass}

As one of the fundamental particles in the universe, massive neutrinos underwent a transition from relativistic to non-relativistic during the evolution of the universe. Low-mass, high-velocity neutrinos suppress the growth of structure at small scales, leaving detectable signatures on the large-scale structure of the universe (\citealt{Lesgourgues_2006}). In this work, we use the Aemulus $\nu$ simulation suite, which includes neutrino mass as a cosmological parameter, to construct our emulators. By combining galaxy clustering and galaxy--galaxy lensing data from BOSS galaxies, we can investigate constraints on neutrino mass within the framework of our model and data.

We first evaluate the sensitivity of our summary statistics to changes in the neutrino mass. In order to do so, we vary the neutrino mass in the emulators while keeping $\Omega_m$ and all other parameters fixed. Different combinations of the galaxy bias parameters from models like HOD can have distinct predictions of the summary statistics at small scales. In order to isolate the coherent behavior of neutrino mass, we repeat the predictions for 400 randomly selected combinations of cosmological and galaxy-–halo connection parameters. Using a reference neutrino mass of 0.011 eV, the lower limit in the simulation suite, we compute the ratio of the summary statistics relative to this baseline.   The results indicate that varying the neutrino mass within the prior range does have some effect on the values of clustering and lensing signals at small scales, but the highest offset for statistics like $w_{p}$ and $\xi_{0}$ is only a few percent, while the effect is weaker for $\xi_{2}$ and $\Delta \Sigma$.

Then we isolate the constraint on neutrino mass with different data combinations. The constraints are dominated by the clustering measurements, with the inclusion of lensing data changing the relative error of neutrino mass by less than ~3.5\%, indicating a minimal impact.  In Appendix \ref{appsec:contours}, we provide the full constraints of all parameters from clustering and lensing measured with HSCz galaxies at three redshifts in our fiducial analysis. Similar results are obtained from combination of clustering and the other two lensing measurement. Overall, the data show a preference for non-zero neutrino mass, but this signal is not quite significant and is largely influenced by the prior distribution and sample variance.

In order to further examine our result, we run additional tests with the neutrino mass fixed at different values. Fixing neutrino mass leads to slight increases in $\chi^2$, but the changes are small compared to the total $\chi^2$ ($\Delta \chi^2 < 1$ for different data combinations and redshifts) and do not indicate a statistically significant preference. We also measure the Bayesian factor from the likelihood analysis and compare the evidence between models with free or fixed value of neutrino mass, and also find that the Bayesian ratio for model comparison is weak. For instance, the ratio between hypothesis with free neutrino mass and fixing neutrino mass to be 0.011 eV is less than 3, implying that adding neutrino mass as a free parameter is not strongly preferred. This also indicates that the current constraint on neutrino mass is not tight enough to fully exclude the massless case as shown by the contour plot. As mentioned earlier, this preference for non-zero neutrino mass could be due to the impact of priors in the likelihood analysis but it can be challenging to quantify. We can adopt a flat prior in the multi-dimensional parameter space but it is inevitable to extrapolate to space without training data such that we can not guarantee the accuracy of the emulators. One indirect solution is to add more information into the analysis so that the prior becomes subdominant and it is worth investigating in future works. 

We note that fixing neutrino mass to a low value such as 0.011 eV is close to the models in Aemulus V for clustering analysis which have no neutrinos. We compare the resultant constraints from clustering only data on the model parameters. Since the priors on the cosmological parameters are slightly different, we find some offsets in parameters of $\Omega_{m}$ and $\sigma_{8}$ at a level of less than $1.2\sigma$. However, due to the degeneracy and the parameter $\gamma_{f}$, the resultant constraint on $f\sigma_{8}$ is quite consistent at a level of less than $\sim0.2-0.5\sigma$ for all three redshift bins, which demonstrates the consistency of our modeling approach across different simulation suites.

\section{Discussion and Conclusion}
\label{sec:conclusion}

In this work, we adopt a simulation-based emulation method to jointly analyze galaxy clustering and galaxy--galaxy lensing at the non-linear scale. By utilizing measurements of galaxies from BOSS, DECaLS, and HSC, we provide tight constraints on both cosmological and galaxy bias parameters. Compared to the small-scale clustering analysis in Z23, the inclusion of lensing data in this work enhances the constraining power, particularly for $S_{8}$, the amplitude of structure in the universe by a factor of more than 2. On the other hand, the combined constraint on  $f\sigma_{8}$ does not show significant improvement, as the primary information on this parameter comes from the velocity field, which the lensing signal is not highly sensitive to. Consequently, the final constraint on $f\sigma_{8}$ remains consistent with the results from our previous work based on clustering-only measurements.

Our overall measurement is generally consistent with some works in the literature that $S_{8}$ is low, as summarized in Figure \ref{fig:S8_constraint}. After accounting for scale truncation and employing a simple empirical model for lensing systematics, the tension on $S_{8}$ is reduced, though full agreement has not yet been achieved. A typical tension of $1\sim2\sigma$ level still remains. On the other hand, recent works, such as \cite{Chen_2024} and \cite{Xu_2024} using various methodologies, and cosmic shear studies with updated photometric redshift calibration methods (\citealt{Wright_2025,Janvry_2025}), indicate that lensing measurements from large-scale structure analysis are in good agreement with CMB, suggesting that it is important to further investigate and revisit these analyses from multiple perspectives. This includes, but is not limited to, refining models of galaxy formation, considering new physics beyond the standard cosmological model, and addressing potential unknown systematics in both modeling and observations.

The Aemulus $\nu$ simulation suite is indeed one of the approaches in this direction. It is well known that massive neutrinos can suppress the growth of structure in a scale-dependent manner. However, our results show that adding neutrino mass as an additional degree of freedom does not resolve the inconsistency between large-scale structure (LSS) analyses and CMB. Although we observe that neutrino mass does have an impact on clustering measurements at small scales, this effect is not strong enough to explain the observed tensions. This can be partially attributed to the fact that both the current sample variance and modeling uncertainties are non-negligible, and shot noise in the measurements may erase some information regarding the neutrino mass. As we can see from the marginalized constraints on neutrino mass in our joint analysis, these constraints are largely dominated by the prior from the simulation suite. Given the current constraints from surveys such as \cite{DESI_2024}, which suggest that the sum of neutrino masses is less than 0.1 eV, the impact on clustering and lensing statistics is minimal. From this perspective, future data with a higher number density of galaxies will play a crucial role in improving these measurements. Additionally, the recent data release from DESI, which covers a broader redshift range and larger volume than SDSS (\citealt{DESI_2024b}), combined with other data products, will enable better measurements of galaxy clustering and galaxy--galaxy lensing. This provides an ideal framework for analyses like the one presented in this work.

An important component of our work is the model describing the connection between galaxies and halos. Since the scales involved in the analysis span both linear and non-linear regimes, it requires prescriptions that incorporate both cosmology and galaxy formation processes. To address this complexity, we have extended our empirical HOD model to include sufficient degrees of freedom, which has been validated against galaxy mocks generated using an independent SHAM model. Based on our current and earlier analyses, our model effectively captures the dominant cosmological information at these scales, particularly the amplitude and growth of structure in the universe. Our results suggest that the model is not significantly affected by modeling systematics.
However, this does not necessarily imply that the current empirical model is ideal for future analyses. Ongoing efforts are focused on using more physically motivated models from hydrodynamical simulations or semi-analytic models (SAMs) for galaxy formation to extend and augment the empirical framework (\citealt{Beltz-Mohrmann_2020, Beltz-Mohrmann_2023, Hadzhiyska_2023, Chapman_2023, Kwan_2023, Kwan_2023b, Ortega-Martinez_2024, Zhai_2025}). These more sophisticated prescriptions will enable deeper investigations into the galaxy--halo connection and galaxy clustering. We will leave such attempts for future work.

This work represents one of the first applications of the Aemulus $\nu$ simulation suite to actual survey data. It provides results consistent with our earlier work and demonstrates the consistencies between different simulation suites, i.e., adding massive neutrinos as an independent particle species does not introduce significant theoretical systematics. Additionally, although our analysis is based only on Tier 2 simulations, the combined simulation boxes from both Tier 1 and Tier 2 will provide a much larger training set for machine learning algorithms and enable wider applications for large scale structure analysis in future studies.

In our current work, we have used the standard two-point correlation function and galaxy--galaxy lensing. However, there is great potential for further improving the cosmological constraints by incorporating additional summary statistics (\citealt{Hahn_2020, LiuWei_2023, Storey-Fisher_2024, Hahn_2024b, Paillas_2024, Massara_2024,Liu_2025}) or performing inference at the field level (\citealt{Jamieson_2023, Stadler_2023}). These approaches will not only enhance the accuracy of the constraints but also minimize model bias, enabling a more precise and robust understanding of both cosmological parameters and galaxy formation processes.

\section*{Acknowledgements}

WG and ZZ are supported by NSFC (12373003), the National Key R\&D Program of China (2023YFA1605600), and acknowledges the generous sponsorship from Yangyang Development Fund. JZ is supported by the National Key Basic Research and Development Program of China (2023YFA1607800 and 2023YFA1607802) and NSFC grants (12573004). This work is also supported by the China Manned Space Program with grant no. CMS-CSST-2025-A04 and No. CMS-CSST-2021-A01. The computations in this paper were run on the $\pi$ 2.0 cluster supported by the Center of High Performance Computing at Shanghai Jiao Tong University and the Gravity Supercomputer at the Department of Astronomy, Shanghai Jiao Tong University.

\software{Python,
NumPy \citep{numpy},
SciPy \citep{scipy},
astropy \citep{astropy},
Matplotlib \citep{matplotlib},
George \citep{george_2014},
Halotools \citep{Halotools},
Corrfunc \citep{Sinha_2020},
MultiNest (\citealt{Feroz_2009, Buchner_2014})
}

\appendix
\section{Recovery test}
\label{appsec:recovery}
\begin{figure}[htbp]
\begin{center}
\includegraphics[width=15.0cm]{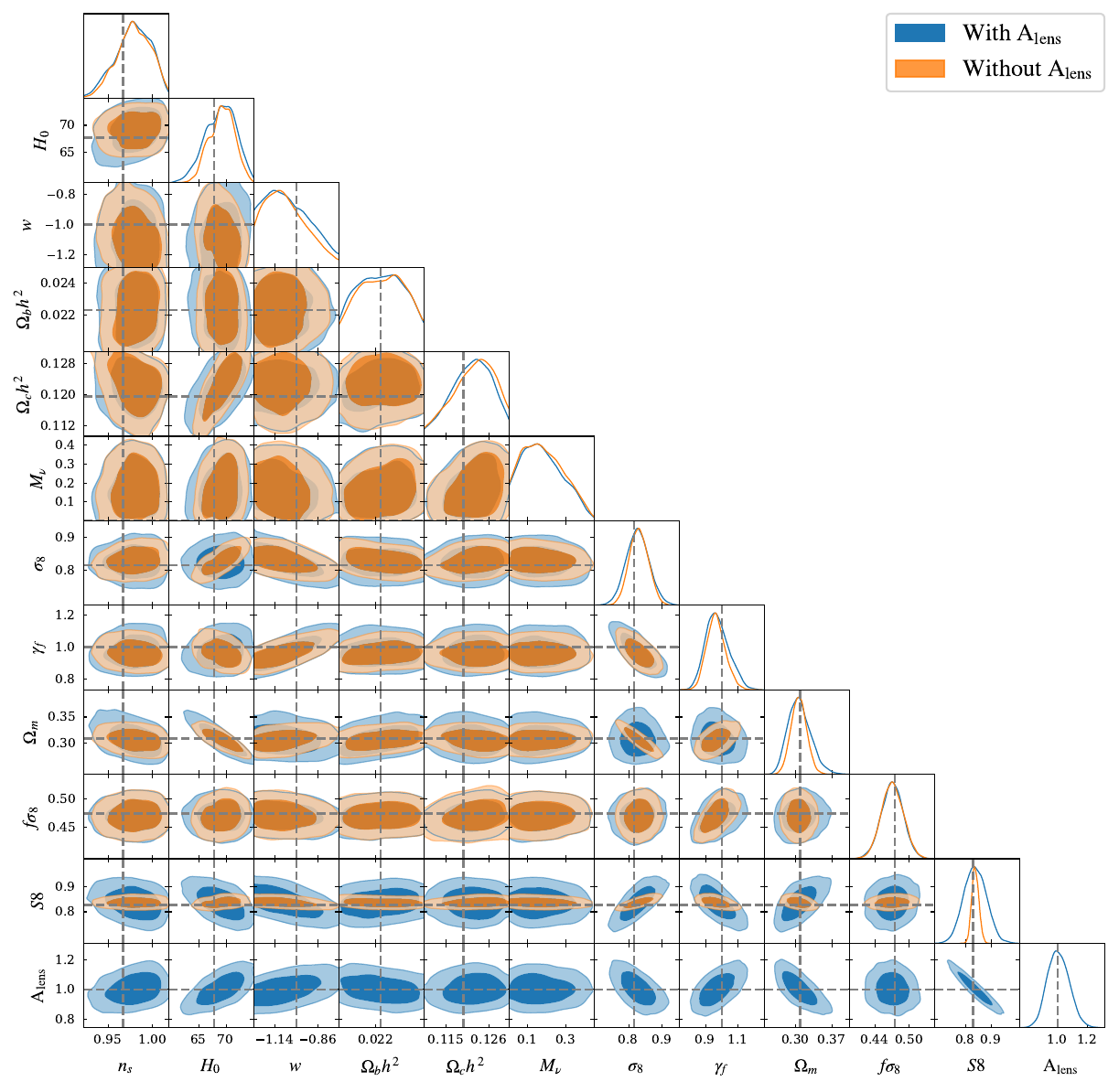}
\caption{Recovery test of the Uchuu SHAM catalog with the dashed lines showing the true values. The result including $\mathrm{A_{lens}}$ is shown in blue, and the result without $\mathrm{A_{lens}}$ is shown in orange.}
\label{fig:constraint_Uchuu}
\end{center}
\end{figure}

In this work, we employ the UNIT and Uchuu simulations, which are at similar redshifts to the high-$z$ data we analyze, to construct the SHAM catalog and perform recovery tests. Our SHAM catalog is based on the method outlined in Z23, which uses a combination of the virial velocity of the halo, $v_{\text{vir}}=(GM_{\text{vir}}/R_{\text{vir}})^{1/2}$, and the maximum
circular velocity within the halo, $v_{\text{max}}$, as the abundance matching parameter (\citealt{Lehmann_2017}). The parameter $\alpha_{\text{sham}}$ is a free parameter that controls the relative importance of these two quantities. When $\alpha_{\text{sham}}=0$, the abundance matching depends only on $v_{\text{vir}}$, meaning that galaxy mass is matched to halo mass. When $\alpha_{\text{sham}}=1$, the abundance matching depends only on $v_{\text{max}}$. Considering the 0 eV neutrino mass of UNIT and Uchuu simulations, in recovery tests we extend the prior lower limit of neutrino mass to 0 eV, which is slightly smaller than in the fiducial likelihood analysis. Figure \ref{fig:constraint_Uchuu} shows the recovery test on the Uchuu SHAM catalog, constructed using $\alpha_{\text{sham}}=1.0$ and scatter=0.15. Dashed lines represent the true values and the 0 eV neutrino mass of Uchuu simulation resides in the boundary. Our model provides reasonable recovery for cosmological parameters in the Uchuu SHAM catalog and all the parameters are within $1\sigma$ of the truth. For the UNIT simulation, we adopt more choices for SHAM parameter $\alpha_{\text{sham}}$ and scatter, and we find similar recovery results compared with Uchuu. Overall, our emulators, based on the HOD model and assembly bias related to the halo environment, deliver robust cosmological constraints.

In order to examine the possible prior-volume effect introduced by $\mathrm{A_{lens}}$, we perform another recovery test on Uchuu SHAM mock, in which we marginalize over $\mathrm{A_{lens}}$. The result including $\mathrm{A_{lens}}$ is shown in red. We find the addition of $\mathrm{A_{lens}}$ yields larger uncertainties for parameter constraints, especially for $S_8$, but the posterior peaks are consistent, indicating that it doesn't induce substantial bias in the constraint.

\section{Performance of emulators} \label{appsec:performance}

\begin{figure}[htbp]
\begin{center}
\includegraphics[width=4.2cm]{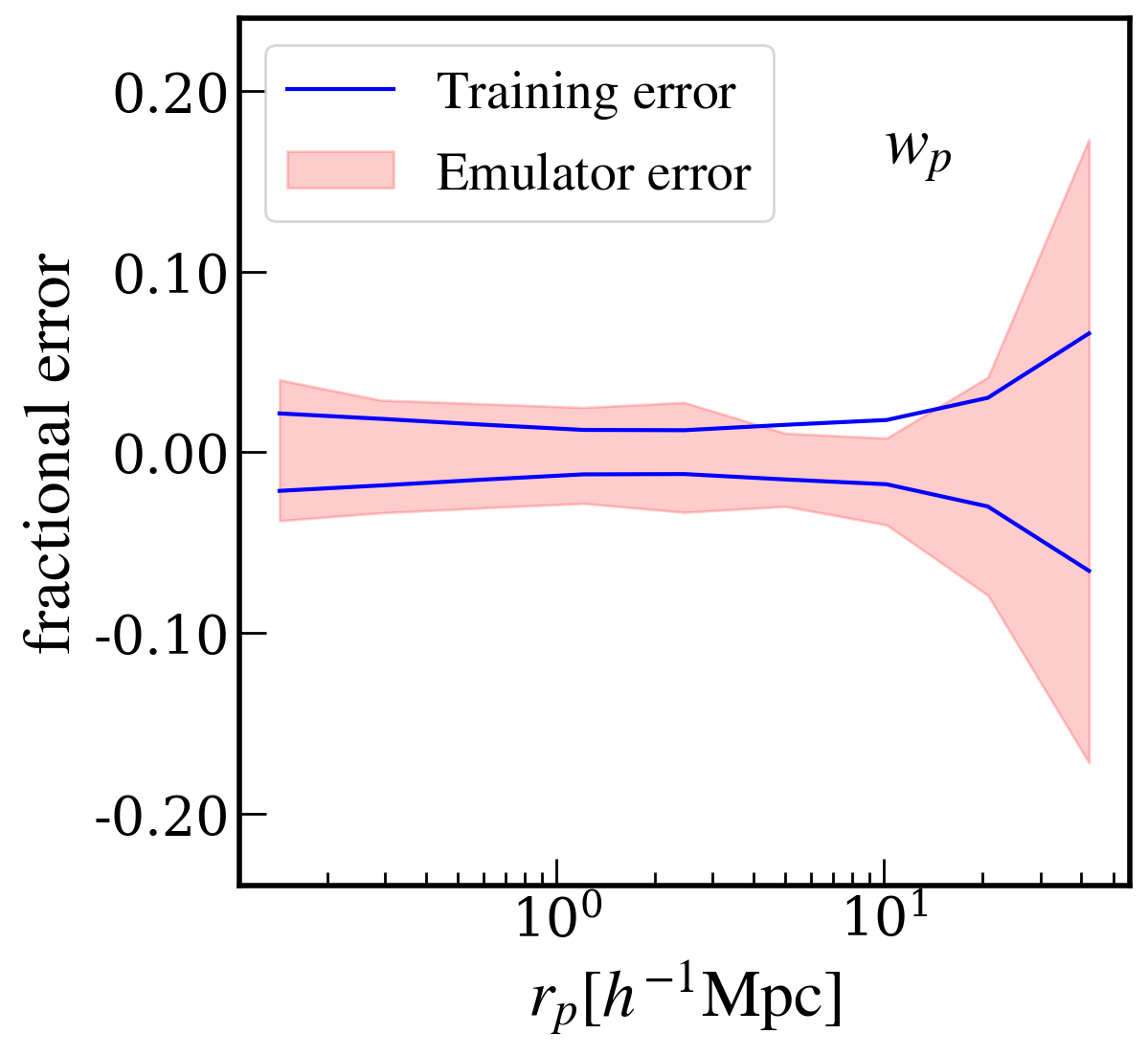}
\includegraphics[width=4.2cm]{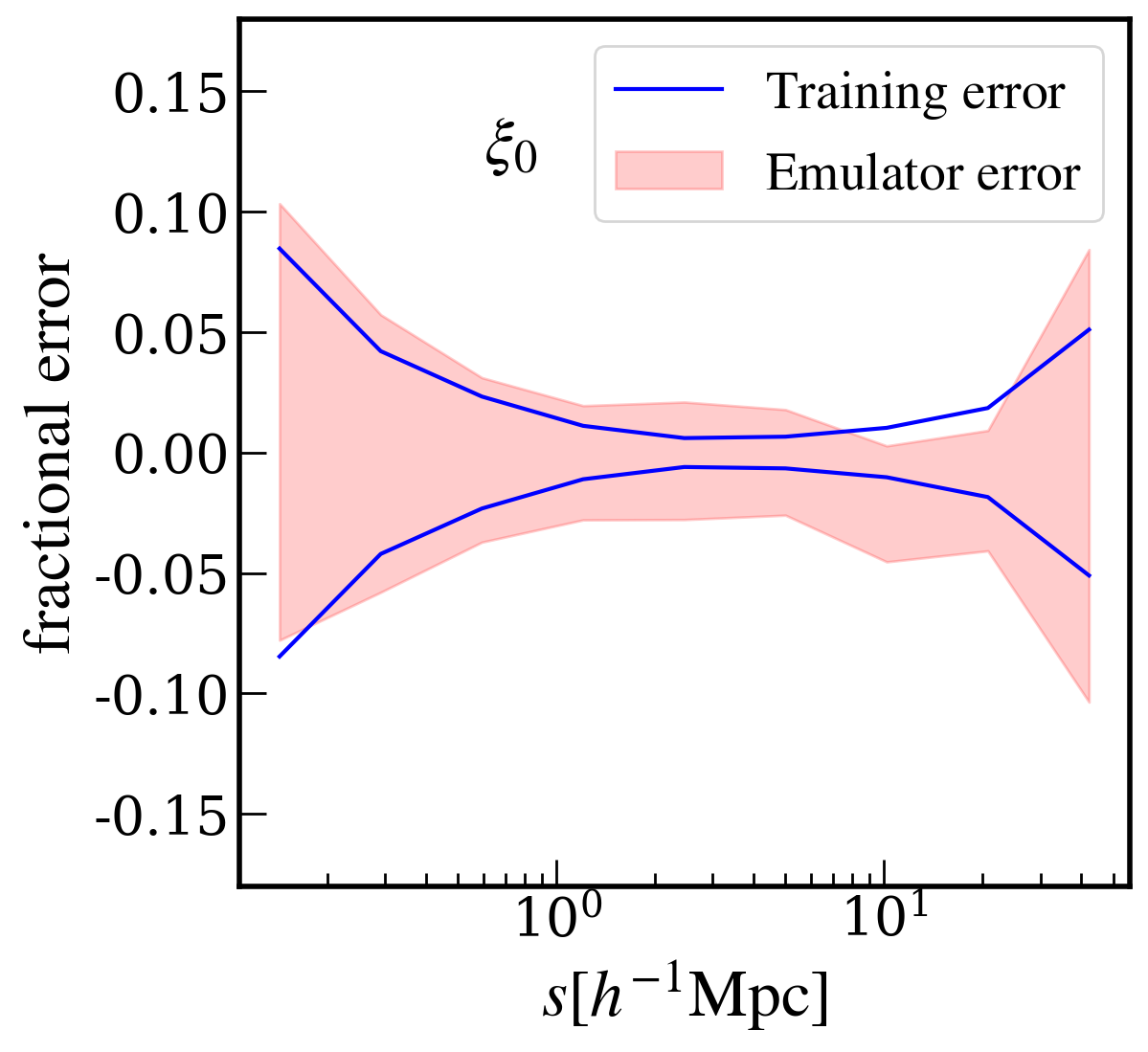}
\includegraphics[width=4.2cm]{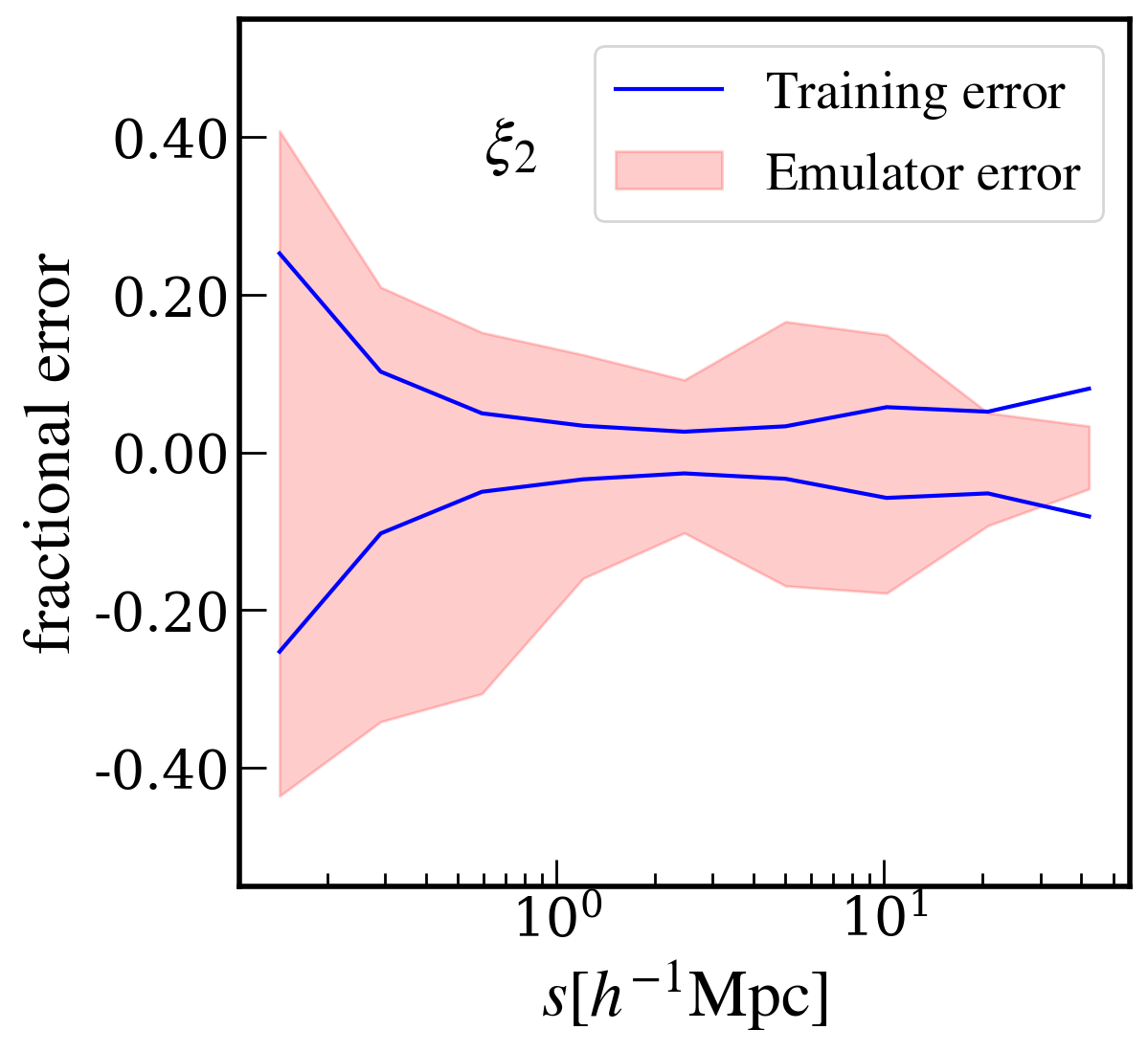}
\includegraphics[width=4.2cm]{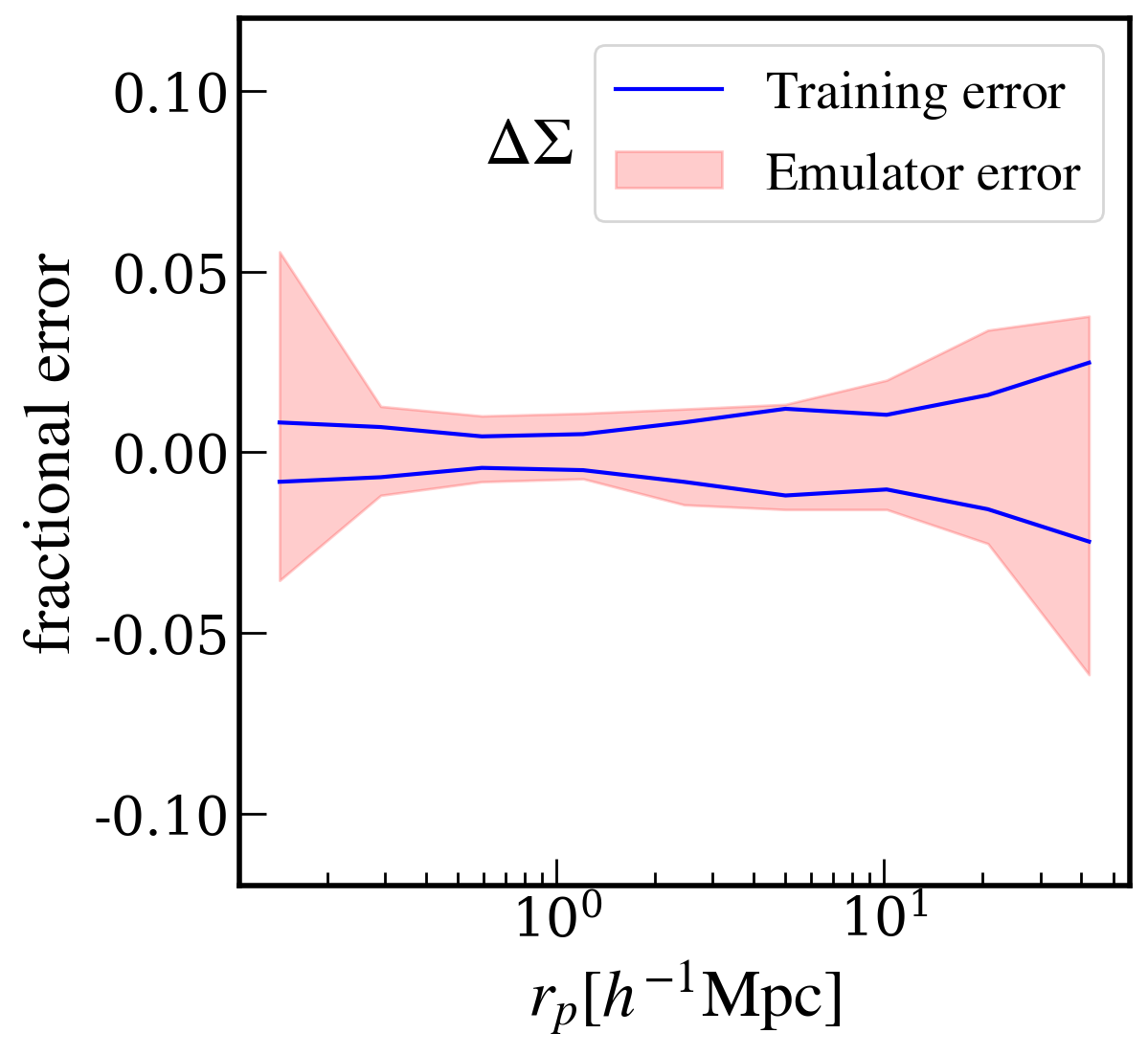}
\caption{Performance of our emulators for different statistics at low-$z$. The blue lines represent the training errors used in the Gaussian Process model, and the red shaded area corresponds to the central 68$\%$ distribution of fraction error for test sets. The performance at the other two redshifts is similar.}
\label{fig:performance}
\end{center}
\end{figure}

In this work, we adopt the methodology from \cite{Zhai_2019} to construct the emulators for summary statistics. We use 40 simulations from the Aemulus $\nu$ Tier 2 boxes as the training set and the remaining 10 simulations as the test set. Each training cosmology has 100 different and non-overlapping HOD models. We use \texttt{Corrfunc} to measure galaxy clustering and employ \texttt{Halotools} to compute galaxy--galaxy lensing from the mock catalogs directly. Figure \ref{fig:performance} shows the fractional error between the values measured from the test set and those predicted by our emulators. For each statistic at different redshifts, we present the central 68$\%$ region of fractional errors for 1000 test parameter sets drawn from the test simulations. Compared to the Aemulus V, we find that the emulators with the new simulation suite for clustering statistics maintains a similar level of accuracy as expected. The galaxy lensing signal generally performs better, reflecting improved accuracy in the emulator predictions. However, we note that the performance of the emulators is affected by the number of particles used to compute the galaxy--galaxy lensing signal, $\Delta \Sigma$. Using more particles reduces shot noise and improves the emulators’ accuracy, but also increases computational time. To balance emulator performance and computational efficiency, we use 1\% of the particles in the simulation (roughly corresponding to 2.7 million particles), which yields reasonable emulator performance.

\section{Test on the Alcock-Paczynski effect}
\label{appsec:ap}
\begin{figure}[htbp]
\begin{center}
\includegraphics[width=8.0cm]{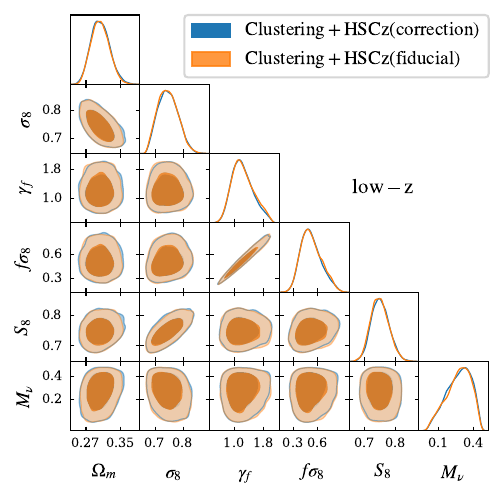}
\caption{Constraints with (blue) and without (orange) AP correction on lensing predictions at low-$z$ using the combination of clustering and lensing measured with HSCz source galaxies.}
\label{fig:correction}
\end{center}
\end{figure}

In our work, we test the potential impact from AP effect correction on lensing predictions, where we perform correction for our lensing measurement from mocks, following \cite{Lange_2023} and \cite{More_2013}. We find the impact on $\Delta \Sigma$ among our cosmology parameters prior is within the range of approximately -3\% to +1\%. Given that the lensing measurements on the scales of interest carry an uncertainty at the level of $\sim$10\%, we regard the effect as subdominant. We further perform likelihood analysis on the observational data using corrected lensing predictions, and find that the resulting posteriors are nearly identical to those without correction. We show the comparison with the lensing measured from HSCz at low redshift in Figure \ref{fig:correction}, with similar results obtained from other datasets and redshifts.

\section{Scale dependence}
\label{appsec:scale}

\begin{figure*}[htbp]
\begin{center}
\includegraphics[width=8.0cm]{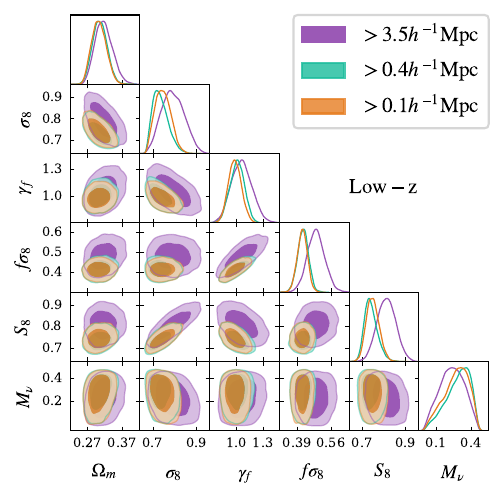}
\includegraphics[width=8.0cm]{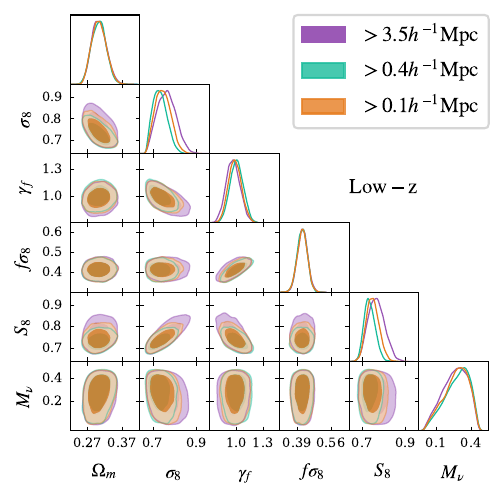}
\caption{Cosmological constraints from different scales for Clustering+HSCz data at low-$z$. The left panel shows constraints considering clustering and lensing data above certain scales, while the right panel shows constraints from the combination of clustering from 0.1 to 60.2 $h^{-1}$Mpc with cut lensing measurement above some scales. The  $>0.4h^{-1}$Mpc and $>3.5h^{-1}$Mpc contours represent constraints above the fiber collision scale and the linear scale, respectively, by excluding data below those scales. The $>0.1h^{-1}$Mpc contours correspond to the situation considering clustering and lensing from 0.1 to 60.2 $h^{-1}$Mpc without any scale cut-off, and are same in both panels.}
\label{fig:scale_constraint}
\end{center}
\end{figure*}

Using a simulation-based emulation method, we extend our analysis to non-linear scales, which contain abundant cosmological information. To evaluate the constraining power from different scales, we exclude statistics below 0.4 $h^{-1}$Mpc and 3.5 $h^{-1}$Mpc and compare the results in the left panel of Figure \ref{fig:scale_constraint}. The 0.4 $h^{-1}$Mpc scale corresponds to the fiber collision scale, and 3.5 $h^{-1}$Mpc represents a scale slightly larger than the transition between the one-halo and two-halo terms. The 0.1 $h^{-1}$Mpc scale corresponds to the scenario considering all statistics from 0.1 to 60.2 $h^{-1}$Mpc. Similar to Z23, we find that ignoring observational data at smaller scales leads to a larger estimate for $f\sigma_8$,  with the effect being more pronounced at lower redshift. Therefore, we present the results at low $z$. In this work, we also observe a similar phenomenon for $S_{8}$. Including highly non-linear scales leads to smaller $f\sigma_8$ and $S_{8}$ values with smaller uncertainties, which increases the tension with Planck. On the other hand, to specifically examine the effect of galaxy--galaxy lensing at different scales, we conduct a similar test where we exclude lensing data below these scales, as shown in the right panel of Figure \ref{fig:scale_constraint}. Excluding lensing at smaller scales mainly leads to slightly larger uncertainties for $S_{8}$, though this effect is less significant than when excluding smaller scale clustering simultaneously. Figure \ref{fig:scale_constraint} shows the results of Clustering+HSCz data and other datasets also show similar results. Thus we conclude that the small scales mostly affect the constraints on clustering, and that the results are consistent enough that we keep the smallest scales for our fiducial analysis. Combining clustering data from 0.1 to 60.2 $h^{-1}$Mpc with lensing data from 1.7 to 60.2 $h^{-1}$Mpc retains most of the constraining power on the main cosmological parameters. 

\section{Galaxy assembly bias}
\label{appsec:bias}
\begin{figure*}[htbp]
\begin{center}
\includegraphics[width=8.0cm]{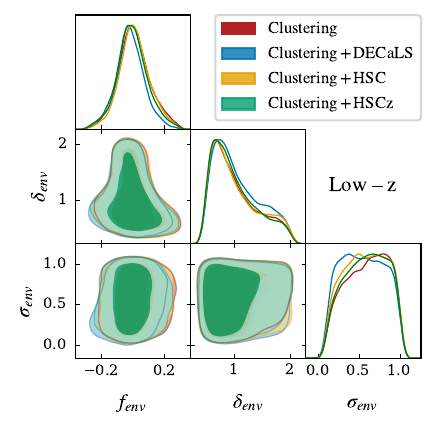}
\caption{Constraints on the three assembly bias parameters for different data at low-$z$. The parameter $f_\text{env}$ measures the dependence on assembly bias.}
\label{fig:ab_constraint}
\end{center}
\end{figure*}

\begin{figure}[htbp]
\begin{center}
\includegraphics[width=8.0cm]{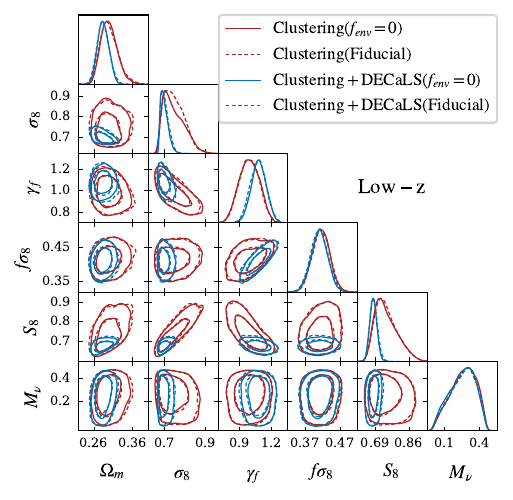}
\includegraphics[width=8.0cm]{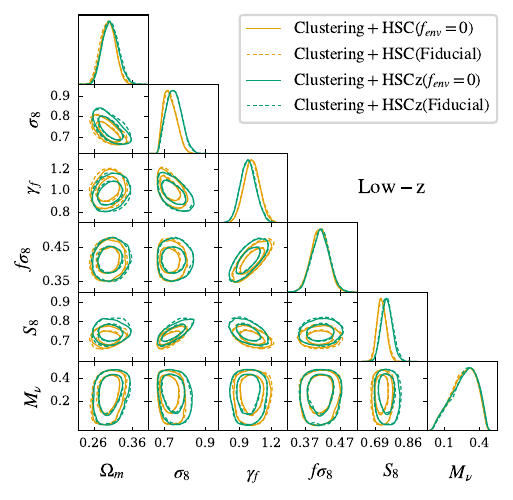}
\caption{Comparison between results without assembly bias by setting $f_\text{env}=0$ and fiducial results at low-$z$. The left panel shows Clustering-only and Clustering+DECaLS results, while the right panel shows Clustering+HSC and Clustering+HSCz results. Solid lines correspond to the $f_\text{env}=0$ case, while dashed lines represent the fiducial results.}
\label{fig:ab_effect}
\end{center}
\end{figure}

Figure \ref{fig:ab_constraint} shows the constraints on three galaxy assembly bias parameters at low redshift. For all four sets of observational data, $f_\text{env}$ is close to 0, indicating that there is minimal assembly bias. To assess the impact of assembly bias on cosmological constraints, we conduct a test in which we fix $f_\text{env}=0$, effectively excluding assembly bias. Figure \ref{fig:ab_effect} presents the constraints on the main cosmological parameters both with and without assembly bias at low redshift. The exclusion of assembly bias has a negligible effect on the cosmological constraints, which is consistent with the fact that $f_\text{env}$ is near 0 in our fiducial analysis. We also find similar results for the other two redshifts.

It is worth noting that this work only considers halo environment as the galaxy assembly bias parameter, which may be an incomplete description in the galaxy--halo connection model. Our results suggest that there is no preference for environment-based assembly bias from BOSS galaxy clustering and galaxy--galaxy lensing data with different source galaxies. Future analyses on observational data with higher accuracy will likely require more comprehensive models of assembly bias (e.g., \citealt{Xu_2021}). A broader set of galaxy assembly bias parameters will be necessary to improve the accuracy and of cosmological constraints( \citealt{Wang_2025}).

\section{Constraint on all parameters}
\label{appsec:contours}

\begin{figure*}[htbp]
\begin{center}
\includegraphics[width=18.0cm]{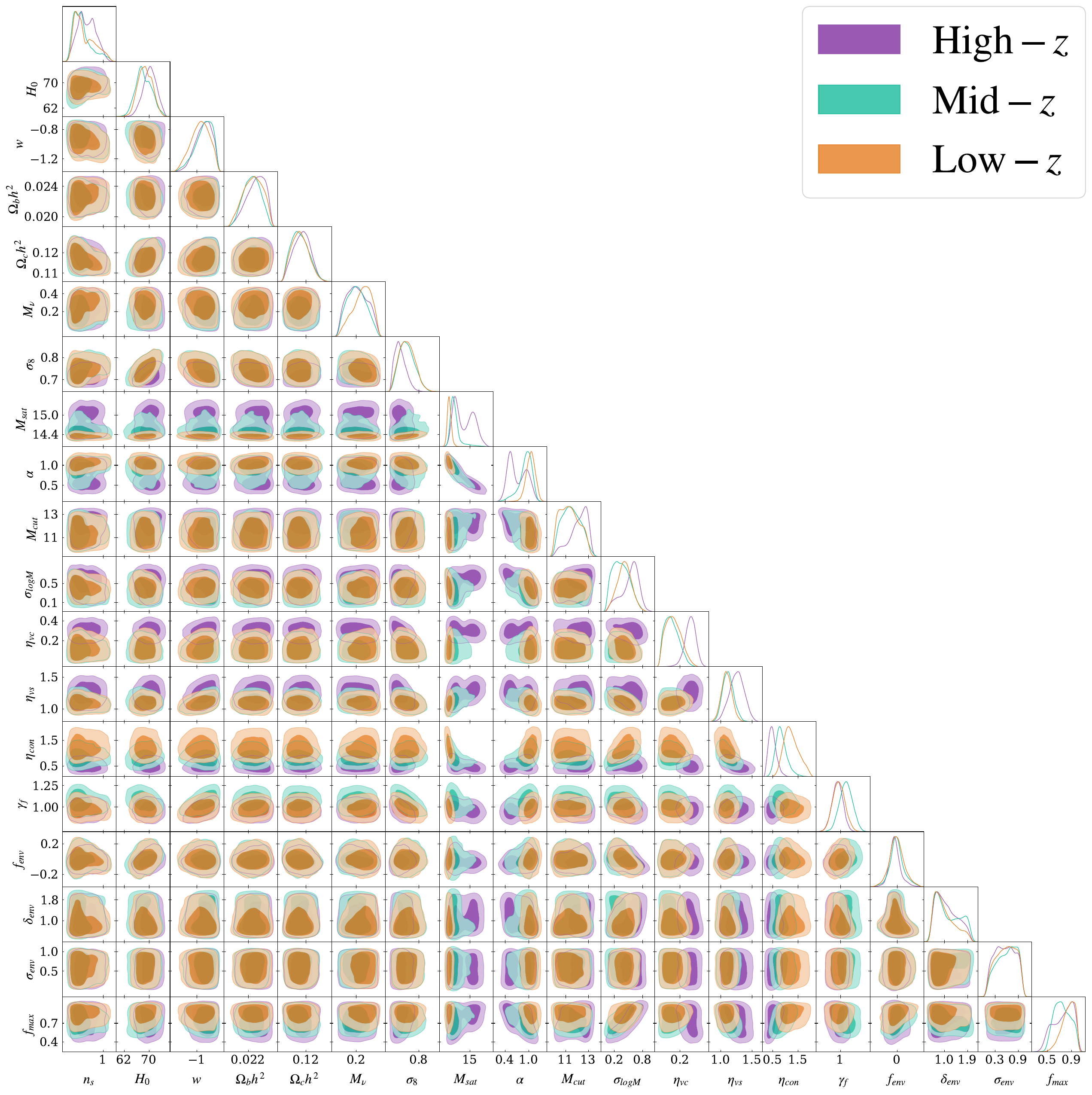}
\caption{Full constraints from clustering and lensing measured with HSCz galaxies for three redshift bins in the fiducial analysis.}
\label{fig:contours}
\end{center}
\end{figure*}

Figure \ref{fig:contours} shows constraints on all parameters from the combination of galaxy clustering and galaxy--galaxy lensing measured with HSCz galaxies across three redshifts in our fiducial analysis. The other combinations of observational data give similar results. In addition, we notice that adding lensing measurements in the analysis can induce a bimodal distribution for a few HOD parameters at high redshift, for instance $M_{\textrm{sat}}$ and $\alpha$, although the other parameters remain consistent. From the scale dependent analysis, we find that this is mainly caused by the offset of fitting the data point of $\xi_{0}$ at the smallest scale. Removing the measurements at this scale can give a clear uni-modal distribution for all the parameters.

\bibliographystyle{yahapj}
\bibliography{emu_gc_bib,software}

\end{document}